\documentclass[useAMS,usenatbib]{mn2e}

\usepackage{amsmath}
\usepackage{amssymb}
\usepackage[dvipsnames]{xcolor}
\usepackage{dirtytalk}
\usepackage{graphicx}
\usepackage{graphics}
\usepackage{hyperref}
\hypersetup{
     colorlinks   = true,
     citecolor    = blue
}

\def\apj{{ ApJ}}
\def\apjl{{ApJL}}

\def\aap{{ A\&A}}

\def\mnras{{ MNRAS}}
\def\araa{{ ARA\&A}}

\def\nat {{ Nature}}

\def\physrep{{ Physics Reports}}

\def\prl{{ Phys. Rev. Lett}}

\def\mr{\mathrm}
\def\d{\mr{d}}
\def\b{\boldsymbol}
\def\t{\widetilde}
\def\mc{\mathcal}
\def\mp{m_{\rm p}}
\def\me{m_{\rm e}}
\def\btheobs{\bar{\theta}_{\rm obs}}
\def\theobs{\theta_{\rm obs}}

\def\epse{\epsilon_{\rm e}}
\def\epsB{\epsilon_{\rm B}}
\def\gm{\gamma_{\rm m}}
\def\xm{x_{\rm m}}
\def\num{\nu_{\rm m}}
\def\gc{\gamma_{\rm c}}
\def\xc{x_{\rm c}}
\def\nuc{\nu_{\rm c}}
\def\sigmaT{\sigma_{\rm T}}

\def\rw{r_{\rm w}}
\def\rdec{r_{\rm dec}}

\def\thec{\theta_{\rm c}}

\def\Mj{M_{\rm j}}
\def\Mji{M_{\mr{j},i}}
\def\sigmaj{\sigma_{\rm j}}
\def\sigmaji{\sigma_{\mr{j},i}}
\def\ib{{i_{\rm b}}}
\def\para{\parallel}
\def\theLOS{\theta_{\rm LOS}}

\newcommand{\myemail}{wenbinlu@caltech.edu}

\title[Relativistic Jet Dynamics and Afterglow]{Deceleration of relativistic
jets with lateral expansion}
\author[Lu, Beniamini \& McDowell]
{Wenbin Lu$^1$\thanks{\myemail}, Paz Beniamini$^1$, and Austin McDowell$^2$\\
  $^1$Theoretical Astrophysics, California Institute of Technology, Mail Code 350-17, Pasadena, CA 91125, USA\\
  $^2$Center for Cosmology and Particle Physics, Physics Department, New York University, New York, NY 10003, USA}

\begin{document}
\label{firstpage}
\maketitle

\begin{abstract}
We present a model for the hydrodynamics of a relativistic jet interacting with the circum-stellar medium (CSM). The shocked CSM and the jet material are assumed to be in an infinitely thin surface, so the original 2D problem is effectively reduced to 1D. From general conservation laws, we derive the equation of motion for each fluid element along this surface, taking into account the deceleration along the surface normal due to newly swept-up mass and lateral expansion due to pressure gradient in the tangential direction. The pressure and energy density of the shocked CSM are given by the jump conditions at the forward shock. The method is implemented with a finite-differencing numerical scheme, along with calculation of synchrotron emission and absorption from shock-accelerated electrons, in a new code $\mathtt{Jedi}$ (for \say{jet dynamics}). We present a number of test cases, including top-hat jet, power-law structured jet, \say{boosted fireball} profile, and CSM with density jump at the wind termination shock. Based on the agreement with other analytical and numerical calculations, we conclude that our simplified method provides a good approximation for the hydrodynamics and afterglow emission for a wide variety of jet structures and CSM density profiles. Efficient modeling of the afterglow from e.g., neutron star mergers, will provide important information on the jet energetics, CSM properties, and the viewing angle.
\end{abstract}

\begin{keywords}
relativistic processes --- hydrodynamics --- methods: numerical --- gamma-ray bursts: general
\end{keywords}

\section{Introduction}
The hydrodynamics of a spherically symmetric blastwave expanding in the surrounding medium were extensively studied across the non-relativistic and relativistic regimes \citep{1950RSPSA.201..159T, 1959sdmm.book.....S, 1976PhFl...19.1130B, 1997ApJ...491L..19W, 1998ApJ...496L...1R,  1999ApJ...512..699C, 1999MNRAS.309..513H, 1999PhR...314..575P, 2006ApJ...651L...1B, 2012ApJ...752L...8P, 2013MNRAS.433.2107N}. The case of a narrowly beamed relativistic jet, such as in gamma-ray bursts (GRBs), is more complex due to the effect of lateral expansion.

Initially, the Lorentz factor $\Gamma$ is likely greater than the inverse of the jet opening angle $\theta_{\rm j}$ and hence the jet dynamical evolution is effectively spherical. As the blastwave decelerates, lateral expansion becomes possible in the causally connected region of the jet. Early analytical models assumed that lateral expansion takes place at the local sound speed in the comoving frame \citep{1999ApJ...525..737R, 1999ApJ...519L..17S}. In the ultra-relativistic limit, the sound speed ($c/\sqrt{3}$) is close to the speed of light, so the jet opening angle is maintained at $\theta_{\rm j}\sim \Gamma^{-1}$ as the jet decelerates. Lateral expansion proceeds exponentially because increasing $\theta_{\rm j}$ leads to even faster deceleration as more circum-stellar medium (CSM) mass is accumulated. This can be seen from energy conservation  $E\propto \theta_{\rm j}^2\Gamma^2 r^3n(r)\sim \mr{constant}$ (for CSM density profile $n(r)$), which gives the evolution of shock radius $r\sim \mr{constant}$. However, 2D (axisymmetric) relativistic hydrodynamic simulations showed that lateral expansion is slower than in the above picture, in the sense that most of the jet energy remains in the initial opening angle until the blastwave slows down to mildly relativistic speeds \citep{2001grba.conf..312G, 2009ApJ...698.1261Z, 2010ApJ...722..235V}, and the flow very gradually approaches spherical symmetry \citep{2012ApJ...751...57D}.

Unfortunately, reliable calculations of the jet dynamics are expensive, especially when a large number simulations are needed to fit observations such as the afterglow from off-axis jet in neutron star merger event GW170817 \citep{2017ApJ...848L..12A, 2018Natur.561..355M, 2018MNRAS.478L..18T, 2018ApJ...856L..18M, 2019Sci...363..968G, 2019ApJ...886L..17H}. Semi-analytical prescriptions for the lateral expansion speed (i.e., the time derivative of the jet opening angle $\theta_{\rm j}$) have been used \citep{2012MNRAS.421..570G, 2018ApJ...865...94D, 2019arXiv190911691R}, but these are designed only for top-hat jets with a sharp edge at $\theta_{\rm j}$ and it is unclear how to apply them for realistic structured jet where the lateral expansion speed varies at different locations.

In this paper, we propose a simple model that captures the main physics of relativistic hydrodynamics in two dimensions. The basic idea is similar to that of \citet{2003ApJ...591.1075K} and \citet{2012ApJ...752L...8P} in that we assume that all the jet and shocked CSM mass is confined in an infinitely thin surface. Thus, our model is effectively one dimensional and the goal is to capture the dynamical evolution of each fluid element on this surface. The pressure and energy density of the shocked CSM are given by the jump condition at the forward shock. Energy conservation determines how the blastwave decelerates along the shock normal (or the fluid velocity vector). Then, pressure gradient in the tangential direction of the surface gives rise to the lateral expansion. In \S\ref{sec:PDE}, we derive the differential equations governing the dynamics. In \S\ref{sec:numeric}, we numerically realize the dynamical evolution using a finite-differencing scheme. Then, a number of test cases are shown in \S\ref{sec:results} and compared to earlier works. Synchrotron emission and absorption from shock-accelerated electrons are calculated in \S\ref{sec:synchrotron}. We summarize and provide a brief discussion of future applications in \S\ref{sec:summary}. We use $t$ for time in the lab-frame (which is at rest with the compact object responsible for jet launching) and $\tau$ for the time in the observer's frame. Additionally, any quantity with a prime $(')$ is measured in the comoving frame of the fluid, and unprimed quantities are for the lab frame.

\section{Jet Dynamical Evolution}\label{sec:PDE}

We start from an axisymmetric jet with angular structure described by an arbitrary function $\d E/\d \Omega(\theta)$, where $\theta$ is the polar angle with respect to the jet axis. The initial four-velocity profile is given by $\b{u}_0(\theta)$, in the radial direction. Hereafter, $\b{u}=\Gamma\beta \hat{\b{v}}$ denotes the spatial components of the four-velocity,  where $\Gamma$ is the Lorentz factor and $\beta = v/c$ is the velocity normalized by the speed of light $c$. The density profile of the CSM $\rho_0(\b{r})$ is an arbitrary axisymmetric function of the position vector $\b{r}$. The initial CSM has negligible velocity before the arrival of the jet-driven shock.

We assume that the jet duration is short (or the radial thickness is small) such that the reverse shock is non-relativistic\footnote{This is known as the ``thin-shell'' case in the GRB literature.} seen in the comoving frame of the unshocked jet. In this case, the pressure and thermal energy of the shocked jet region are negligible compared to the bulk kinetic energy. If this assumption breaks down, our model still applies at sufficiently late time after the reverse shock has crossed the entire jet and the thermal energy of the shocked jet region has diminished as a result of adiabatic losses.

We assume that, at any moment, both the jet material and the shocked CSM are located in an infinitely thin surface (hereafter ``the jet surface''). At position $\b{r}$ on the surface, the mass column densities for the jet material and the swept-up CSM are $\sigma_{\rm j}(\b{r})$ and $\sigma(\b{r})$ respectively, and the local four-velocity is $\b{u}(\b{r})$. In the following, we present a model for the dynamical evolution of these quantities, under mass, energy, and momentum conservation laws.

The pressure, rest-mass density, and energy density in the comoving frame of the shocked CSM region are given by the jump conditions at the forward shock 
\begin{equation}
  \label{eq:1}
  P'(\b{r}) = {4\over 3}(\Gamma^2-1)\rho_0(\b{r}) c^2,
\end{equation}
\begin{equation}
  \label{eq:6}
  \rho'(\b{r}) = 4\Gamma \rho_0(\b{r}),
\end{equation}
\begin{equation}
  \label{eq:4}
  e'(\b{r}) = 4\Gamma^2 \rho_0(\b{r})c^2,
\end{equation}
where we have taken an equation of state with adiabatic index of $(4+\Gamma^{-1})/3$ \citep[see e.g.,][]{2012ApJ...761..147U}. Under the thin shell assumption, we ignore the pressure inhomogeneity of the shocked CSM in the direction normal to the shock surface. This may cause significant error near the transition region where the CSM density profile changes rapidly, because in reality, it takes a sound-crossing time for the updated pressure information near the shock front to propagate through the shell of shocked CSM. However, the pressure gradient perpendicular to the shock normal must be taken into account to capture lateral expansion. At a given moment, the velocity vector $\b{u}(\b{r})$ is along the local shock normal, so the hydrodynamic force due to transverse pressure gradient $\nabla_\perp P'$ will bend the fluid's trajectory by adding a velocity component along the surface tangent direction. This causes the jet to expand laterally.

Consider a small segment of the jet surface of area $\delta A$ near position $\b{r}$. The rest masses of the jet material and swept-up CSM are given by $M_{\rm j} = \sigma_{\rm j}\delta A$ and $M = \sigma \delta A$. The total energy of this fluid element is given by $E = \Gamma \Mj c^2 + T^{00} V$, where $V$ is the volume and $T^{00} = (e' + P')\Gamma^2 - P'$ is the ``00'' component of the relativistic energy-momentum tensor in the lab frame. Making use of eqs. (\ref{eq:1}--\ref{eq:4}) and $M = \Gamma \rho' V$, we obtain
\begin{equation}
  \label{eq:5}
  E = \Gamma \Mj c^2 + \Gamma^2(1 + \beta^4/3) Mc^2.
\end{equation}
After a short time step $\d t$, this fluid element will sweep up additional amount of rest mass $\d M = \delta A \beta c \rho_0(\b{r})\, \d t$ and decelerate to Lorentz factor $\t{\Gamma} = \Gamma + \d \Gamma$ ($\d \Gamma$ generally being negative). From energy conservation, we obtain
\begin{equation}
  \label{eq:7}
  E + \d M c^2 = \t{\Gamma} \Mj c^2 + \t{\Gamma}^2(1 + \t{\beta}^4/3)
  (M + \d M)c^2.
\end{equation}
Combining eqs. (\ref{eq:5}) and (\ref{eq:7}) and only retaining linear-order terms, we obtain
\begin{equation}
  \label{eq:8}
  \begin{split}
    \d \Gamma &= - {4(\Gamma^2 - 1)\beta^2 \over 3\Mj/M +
    2(4\Gamma - 1/\Gamma^3)} {\d M\over M},\\
  & = {4(\Gamma^2 - 1)\beta^2 \over 3\sigmaj/\sigma +
    2(4\Gamma - 1/\Gamma^3)} {\beta c \rho_0(\b{r}) \over \sigma} \d t
  \end{split}
\end{equation}
For the spherically symmetric case, the above equation describes the deceleration of a thin shell propagating through the ambient medium, consistent with the ultra-relativistic \citep{1976PhFl...19.1130B} and non-relativistic \citep{1950RSPSA.201..159T, 1959sdmm.book.....S} limits. This was recently discussed by  \citet{2012ApJ...752L...8P}. We note that \citet{2012ApJ...752L...8P} did not take into account a (small) derivative term of the adiabatic index with respect to the Lorentz factor and hence their evolution does not strictly conserve energy (causing $\sim$10\% violation during the Newtonian transition). It should also be pointed out that our eq. (\ref{eq:8}) does not properly treat adiabatic loss as the thermodynamic history is not self-consistently included. In particular, when the CSM density profile is $\rho_0\propto r^{-3}$ or steeper, our treatment does not capture the re-acceleration driven by PdV work. A better but more sophisticated solution is proposed by \citet{2013MNRAS.433.2107N}.

In the tangent direction to the jet surface, pressure gradient adds a velocity kick perpendicular to the local velocity vector. Momentum conservation in the comoving frame gives
\begin{equation}
  \label{eq:9}
  {1\over c^2} {\d v_\perp' \over \d t'} = -{\nabla_\perp P'
  \over e' + P' + \rho' c^2 \sigmaj/\sigma},
\end{equation}
where $\nabla_\perp P' = \d P'/|\d \b{r}|$ (since $|\d \b{r}'| = |\d \b{r}|$) and we have assumed that the jet material is simply ``dragged along'' when the shocked CSM spreads in the lateral direction. In reality, the shear at the contact discontinuity between the jet material and the shocked CSM is subject to Kelvin–Helmholtz instability. The amount of mixing and the resulting viscosity is uncertain. However, our prescription is accurate in both limits of $\Mj \gg \Gamma M$ and $\Mj \ll \Gamma M$, because lateral expansion is limited by causality before the jet decelerates significantly ($\Mj \gg \Gamma M$) and the inertia of the jet material is subdominant after significant deceleration ($\Mj \ll \Gamma M$). Making use of $\d t' = \d t/\Gamma$, we obtain a more convenient expression than eq. (\ref{eq:9}) for later numerical implementation
\begin{equation}
  \label{eq:10}
  \d v_\perp' = - \left(4\Gamma^2 \rho_0(\b{r}) c\right)^{-1}
  \left[{\sigmaj\over \sigma} +
    {4\Gamma^2-1\over 3\Gamma}\right]^{-1} {\d P'\over |\d \b{r}|} \d t.
\end{equation}

The qualitative difference between our treatment of lateral expansion and the simplistic local sound-speed prescription of \citet{1999ApJ...525..737R} can be understood by comparing our $|v_\perp'|/c$ to $1/\sqrt{3}\sim 1$. Long after the deceleration time but when the jet is still highly relativistic, eq. (\ref{eq:9}) roughly gives
\begin{equation}\label{eq:LE}
    {v_\perp'\over c} \sim -ct' {\nabla_\perp P'\over 4P'} \sim -{1\over 4\Gamma P'} {\d P'\over \d \theta}.
\end{equation}
The pressure profile is rather flat within the jet core $\theta\ll \thec$, where $\thec$ being the angular size of the jet core. We generally have $\d P'/\d \theta \sim (\theta/\thec)^a P'/\thec$ with $a>0$ and hence $|v_\perp'|/c\sim (\Gamma\thec)^{-1} (\theta/\thec)^a$. At angles $\theta\gg \thec$ in the jet wing region, the pressure profile may be described by a power-law, so $\d P'/\d \theta \sim P'/\theta$ and hence $|v_\perp'|/c\sim (\Gamma\theta)^{-1}$. The jet break occurs when $\Gamma\thec \simeq 1$, and at this time the lateral expansion speed $v_\perp'/c$ is much smaller than unity (as in the sound-speed prescription) by a factor of $(\theta/\thec)^a$ within the jet core and a factor of $\thec/\theta$ in the jet wing. This qualitative difference arises because lateral expansion is driven by smoother pressure \textit{gradient} rather than a rarefaction wave expanding into vacuum \citep[see also][]{2003ApJ...591.1075K, 2012MNRAS.421..522L, 2012MNRAS.421..570G}.

We denote the tangent vector of the jet surface as $\hat{\b{e}}_{\perp} (\b{r})$, since it is perpendicular to the local velocity vector. It can be shown by Lorentz transformation that the change in  four-velocity caused by the transverse kick is $\d \b{u}_\perp = \d v_\perp' \hat{\b{e}}_{\perp}$. Since $\d \b{u}_\perp$ only contributes a second-order change in Lorentz factor (or kinetic energy), energy conservation (eq. \ref{eq:8}) is unaffected by the transverse kick. Thus, the deceleration parallel to the velocity vector is given by $\d \b{u}_\para = \d (\Gamma \beta) \hat{\b{u}} = \beta^{-1}\d \Gamma \hat{\b{u}}$.

\begin{figure}
  \centering
\includegraphics[width = 0.4\textwidth,
height=0.22\textheight]{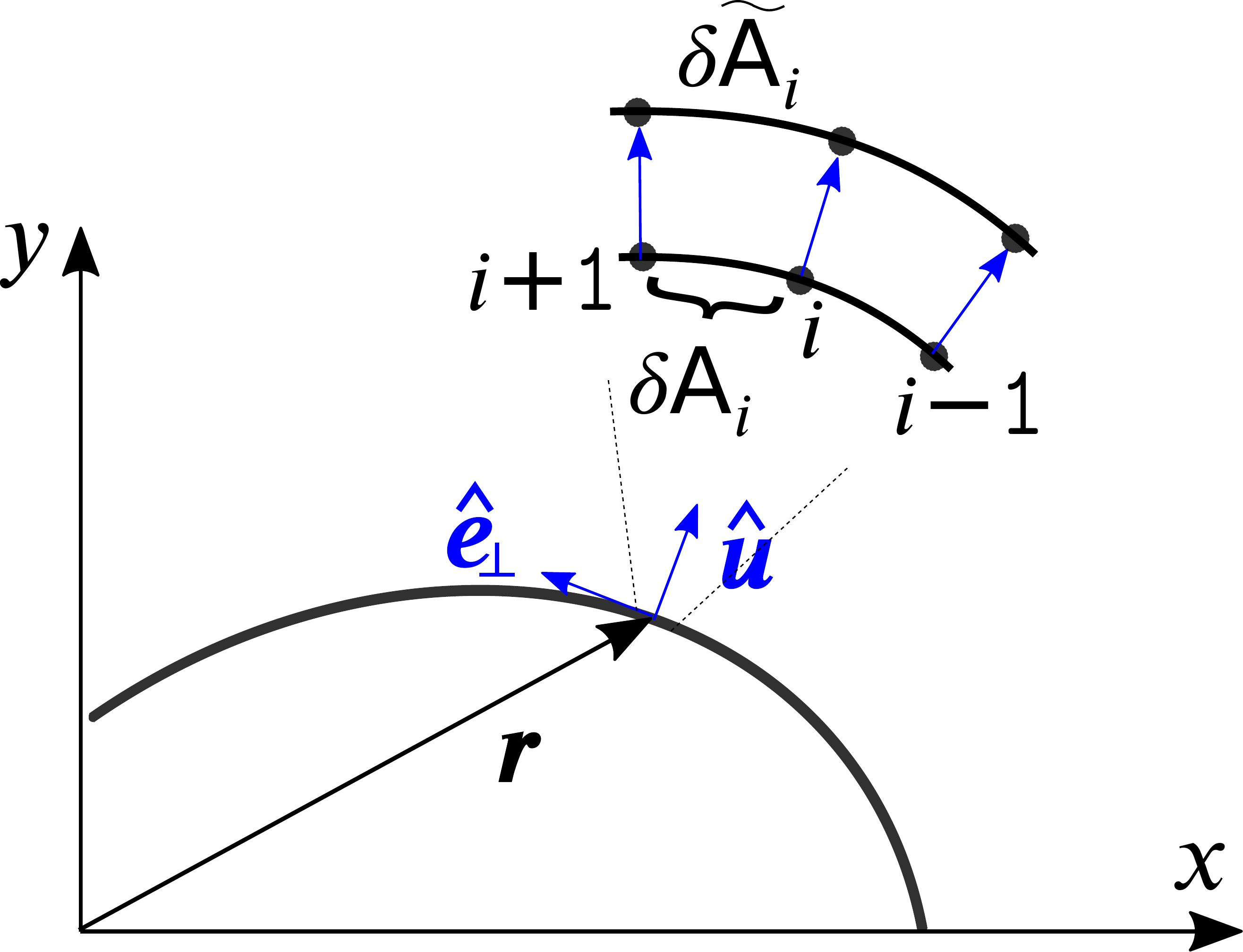} 
\caption{Sketch of the model. The jet axis is along the $\hat{\b{x}}$ direction. The jet material and the shocked CSM are confined in a thin surface shown as the thick black solid line. The surface is numerically discretized into Lagrangian grid points $\{\b{r}_i\}$. The velocity vector $\hat{\b{u}}$ of each point on  the surface is along the local surface normal. The dynamical evolution of each grid point is determined by deceleration caused by newly swept-up CSM and bending of the velocity vector due to pressure gradient in the tangential direction.
}\label{fig:coord}
\end{figure}

Therefore, after each time step $\d t$, we update the four-velocity vector of each grid point by
\begin{equation}
  \label{eq:11}
  \d \b{u} = {\d \Gamma \over \beta} \hat{\b{u}} + \d v_\perp'
  \hat{\b{e}}_\perp,
\end{equation}
and update the position by
\begin{equation}
  \label{eq:12}
  \d \b{r} = \beta \d t \, \hat{\b{u}},
\end{equation}
where $\d \Gamma$ is given by energy conservation (eq. \ref{eq:8}) and $\d v_\perp'$ is given by momentum conservation in the transverse direction (eq. \ref{eq:10}). The mass column densities vary by
\begin{equation}
  \label{eq:13}
     \d \sigmaj  = \sigmaj \left({\delta A\over \delta \t{A}} - 1\right),
\end{equation}
\begin{equation}
  \label{eq:14}
   \d \sigma = \sigma \left[\left(1 + {\beta c \rho_0 \over
       \sigma} \d t\right) {\delta A\over \delta \t{A}} - 1\right],
\end{equation}
where $\delta \t{A}$ is the surface area at $t + \d t $, given by the updated locations of the boundary grids. 

To summarize, we have obtained a dynamical model for the interaction between a structured jet and the CSM with arbitrary density profile. The advantage of our model, over two-dimensional relativistic hydrodynamic simulations, is that the system is effectively one-dimensional with discretization only along the jet surface.

\section{Numerical Method}\label{sec:numeric}
We first describe the initial conditions, then present the numerical scheme based on the finite differencing, and finally discuss the boundary conditions.

At the initial time $t=t_0$, we discretize the jet surface linearly in polar angles $\theta\in [0, \pi/2)$ into $N$ grids according to the initial conditions of $\d E/\d \Omega(\theta)$ and $\b{u}_0(\theta)$. Note that $\d E/\d \Omega$ includes the rest mass energy of the jet material. The inner-most grid point is at $\theta=0$. The location and velocity of each grid point are
\begin{equation}
  \label{eq:2}
  \b{r} = \beta_0 c t_0 \hat{\b{r}},\ \ \b{\beta}_0 = (u_0/\gamma_0)\hat{\b{r}},\ \
  \gamma_0 = (u_0^2 + 1)^{1/2}.
\end{equation}
The initial column densities of jet material and swept-up CSM are
\begin{equation}
  \label{eq:15}
  \sigmaj = {\d E/\d \Omega \over \gamma_0 r_0^2 c^2},\ \ \sigma =
  {1\over 3}\rho_0(\b{r}) r.
\end{equation}
More realistically, the swept-up CSM column density should be an integral along the radial direction from the origin to the current location, but the difference is negligible as long as we choose an initial time $t_0$ much smaller than the deceleration time.

The goal is to trace the motion of each grid point $\{\b{r}_i\}$, $\{\b{u}_i\}$ in a Lagrangian manner, and calculate the evolution of column densities $\sigma_i \equiv \sigma(\b{r}_i)$ and $\sigma_{\mr{j},i} \equiv \sigmaj(\b{r}_i)$ at the grid points.

The surface area of the $i$-th surface segment between $\b{r}_i$ and $\b{r}_{i+1}$ is denoted as $\delta A_i$, which is given by
\begin{equation}
    \delta A_i = \pi (y_i + y_{i+1}) |\b{r}_{i+1} - \b{r}_i|.
\end{equation}
The jet and CSM masses within this surface element of $\delta A_i$ are $M_{\mr{j},i}$ and $M_i$. After a time step $\d t$, we update the grid positions by
\begin{equation}
    \d \b{r}_i = \beta_i c \hat{\b{u}_i}\, \d t,
\end{equation}
and then the updated surface area is
\begin{equation}
    \delta \t{A}_i = \pi (y_i + \d y_i + y_{i+1} + \d y_{i+1}) |\b{r}_{i+1} + \d \b{r}_{i+1} - \b{r}_i - \d \b{r}_i|.
\end{equation}
The amount of newly swept-up CSM mass by the $i$-th surface element is
\begin{equation}
    \d M_i = {(\delta A_i + \delta \t{A}_i)(|\d \b{r}_i| + |\d \b{r}_{i+1}|)\over 4} \cdot \rho_0(\b{r}_i + \d \b{r}_i/2),
\end{equation}
whereas $M_{\mr{j},i}$ stays unchanged. The column densities on the grid points are given by averaging between the two neighboring surface elements, so the change in $\sigma_i$ and $\sigma_{\mr{j},i}$ after $\d t$ are given by
\begin{equation}
    \d \sigma_i = {1\over 2}\left({M_i + \d M_i\over \t{A}_i} + {M_{i-1} + \d M_{i-1}\over \t{A}_{i-1}} - 
        {M_i \over A_i} - {M_{i-1} \over A_{i-1}}\right),
\end{equation}
and
\begin{equation}
    \d \sigma_{\mr{j},i} = {1\over 2}\left( {\Mji\over \t{A}_i} +  {M_{\mr{j}, i-1}\over \t{A}_{i-1}} -
        {\Mji \over A_i} -{ M_{\mr{j}, i-1} \over A_{i-1}}\right),
\end{equation}
Note that the column density $\sigma_i$ on the grid point at $\b{r}_i$ is different from $M_i/A_i$ (which is the column density at position $\b{r}_i + |\b{r}_{i+1}-\b{r}_i|/2$). The first-order difference is important for maintaining numerical stability.

The pressure gradient at the $i$-th grid point is given by
\begin{equation}
  \label{eq:39}
  \left({\d P'\over |\d \b{r}|}\right)_i = {1\over 2}\left({P'_{i+1} - P'_{i} \over
    |\b{r}_{i+1} - \b{r}_{i}|} + {P'_{i} - P'_{i-1} \over
    |\b{r}_{i} - \b{r}_{i-1}|}\right).
\end{equation}
This leads to the tangential velocity kick
\begin{equation}
\label{eq:vperp_i}
    \d v_{\perp,i}' = - {\d t \over 4\Gamma_i^2 c \rho_0(\b{r}_i + \d \b{r}_{i}/2)}
  \left[{\sigmaji\over \sigma_i} +
    {4\Gamma_i^2-1\over 3\Gamma_i}\right]^{-1} \left({\d P'\over |\d \b{r}|}\right)_i.
\end{equation}
The change in Lorentz factor is
\begin{equation}
    \d \Gamma_i = - {2(\Gamma_i^2 - 1)\beta_i^2 \over 3\sigmaji/\sigma_i + 2(4\Gamma_i - 1/\Gamma_i^3)}
    \left({\d M_i\over M_i} + {\d M_{i-1} \over M_{i-1}} \right).
\end{equation}
Therefore, we update the velocity of the $i$-th grid point by
\begin{equation}
    \d \b{u}_i = {\d \Gamma_i\over \beta_i} \hat{\b{u}}_i + \d v_{\perp,i}' (\hat{\b{z}}\times \hat{\b{u}}_i). 
\end{equation}

Finally, we describe our choice of boundary conditions. We place the 0-th grid point at $\theta=0$ (on the jet axis), so it does not experience any lateral expansion ($\d v_\perp'=0$) and the velocity $\b{u}_0$ is always along the $\hat{\b{x}}$ direction. Note that here the subscript $_0$ (meaning the 0-th grid) is not to be confused with that of the initial conditions. The evolution of the column densities are taken as
\begin{equation}
    \d \sigma_0 = {M_0 + \d M_0 \over \t{A}_0} - {M_0\over A_0},\ \ \d \sigma_{\mr{j},0} = {M_{\mr{j},0}\over \t{A}_0} - {M_{\mr{j},0} \over A_0}.
\end{equation}
The evolution of Lorentz factor and velocity are taken as
\begin{equation}
    \d \Gamma_0 = - {4(\Gamma_0^2 - 1)\beta_0^2 \over 3\sigma_{\mr{j},0}/\sigma_0 + 2(4\Gamma_0 - 1/\Gamma_0^3)} {\d M_0\over M_0},\ 
    \d u_0 = {\d \Gamma_0\over \beta_0}.
\end{equation}

We use outflow boundary conditions for the outer boundary at $\theta= \pi/2$, i.e., a grid point is removed once it moves past the equatorial plane\footnote{In reality, due to the existence of a counter jet, a shock forms when the outer wing of the jet reaches the equatorial plane. We ignore the consequence of this shock since only a small fraction of the total energy is involved.}. For the $i_{\rm b}$-th grid point that is closest to the outer boundary at $\theta = \pi/2$, the column density evolution is taken as the linear extrapolation from inner grids
\begin{equation}
    \d \sigma_{\ib} = \d \sigma_{\ib-1} + |\b{r}_{\ib} - \b{r}_{\ib-1}| {\d \sigma_{\ib-1} - \d \sigma_{\ib-2} \over |\b{r}_{\ib-1} - \b{r}_{\ib-2}|},
\end{equation}
and similarly for $\d \sigma_{\mr{j},\ib}$. The ratio $\d M_{\ib}/M_{\ib}$ involved in the Lorentz factor evolution $\d \Gamma_{\ib}$ is obtained using the same extrapolation. As for the pressure gradient, we use the "upwind" prescription
\begin{equation}
  \label{eq:40}
  \left({\d P'\over |\d \b{r}|}\right)_{i_{\rm b}} = {P'_{i} - P'_{i-1} \over |\b{r}_{i} - \b{r}_{i-1}|},
\end{equation}
which then gives the tangential velocity kick from eq. (\ref{eq:vperp_i}).

The above numerical scheme has been implemented in the code $\mathtt{Jedi}$ (for \say{jet dynamics}) in C++. Test runs give stable evolution from highly relativistic initial conditions (e.g., $\Gamma\sim 10^3$) to Newtonian speeds (e.g., $\beta\sim 0.1$) in a few seconds on a 2.3 GHz Intel core. Total energy is conserved to better than $1\%$. Before presenting the results in the next section, we discuss the machine units used in the numerical calculations.

Physical quantities are converted into dimensionless numbers using the deceleration radius/time of a spherically symmetric blastwave of energy $E_{\rm iso}$ and four-velocity $u_{\rm 0,max}$ in a medium of constant density $\rho_{\rm norm}$, as follows
\begin{equation}
  \label{eq:44}
  \begin{split}
    \mbox{length unit: }&  r_{\rm dec} = \left(3E_{\rm iso} \over 4\pi
      u_{\rm 0,max}^2 \rho_{\rm norm} c^2\right)^{1/3},\\
    \mbox{time unit: }& t_{\rm dec} = r_{\rm dec}/c,\\ 
    \mbox{mass unit: }& E_{\rm iso}/c^2,
  \end{split}
\end{equation}
where $E_{\rm iso} \equiv 4\pi \d E/\d \Omega (\theta=0)$, $u_{\rm 0,max} \equiv u_0(\theta=0)$, $\rho_{\rm norm}$ is a normalization constant in the CSM density function $\rho_0(\b{r})$. The above units conversion is equivalent to normalizing the jet energy structure such that the peak isotropic energy is unity, and normalizing the spacetime such that the deceleration radius and deceleration time are both unity. In this way, one single scale-invariant simulation represents an entire family of physical cases. The same scalings have been used by the afterglow-fitting code $\mathtt{Boxfit}$ \citep{2012ApJ...749...44V} to reduce the number of numerical runs.

\begin{figure}
  \centering
\includegraphics[width = 0.45\textwidth, keepaspectratio]{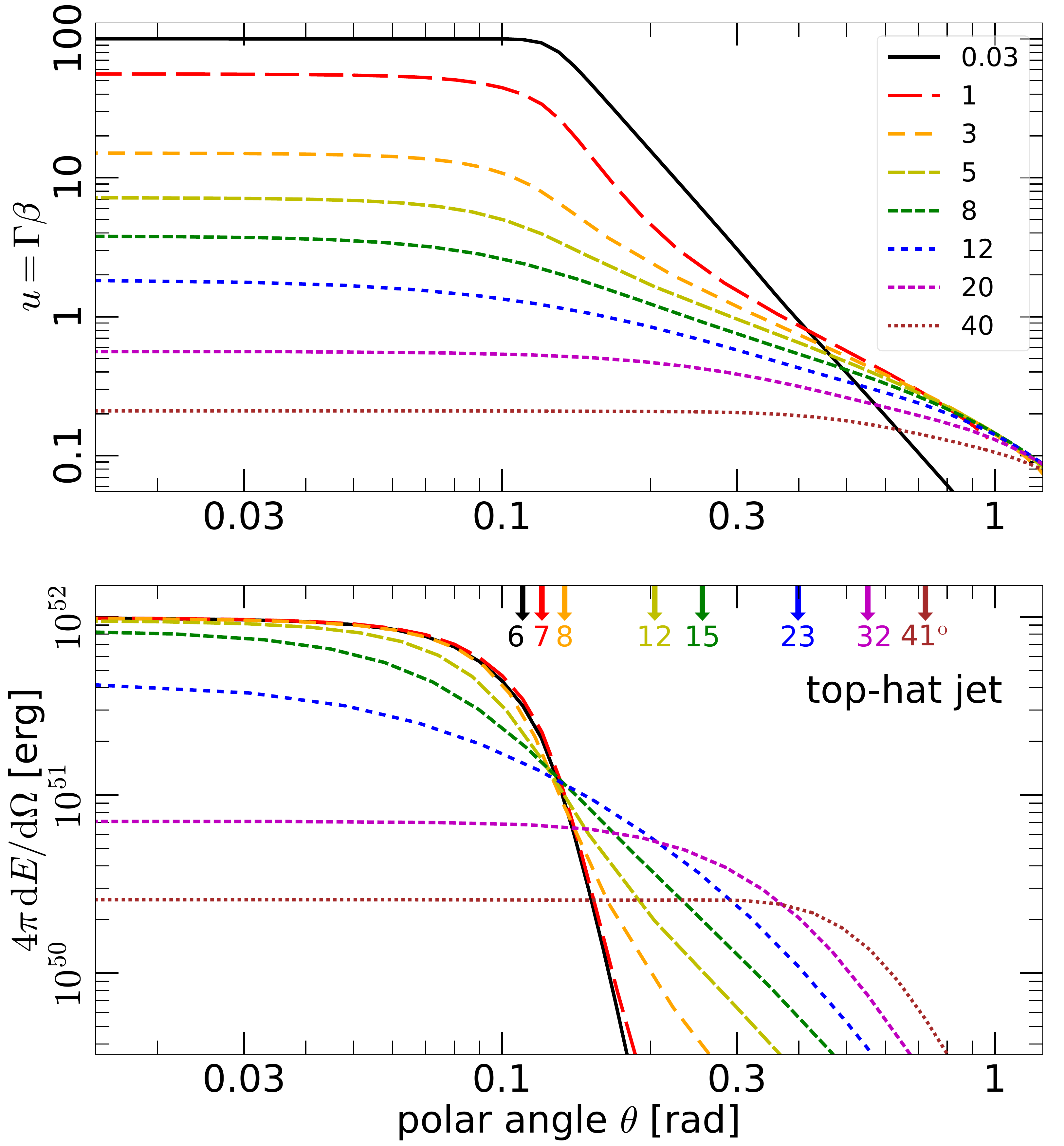}
\caption{Evolution of angular structures $u(\theta)$ (upper panel) and $\d E/\d \Omega(\theta)$ (lower panel) for a top-hat jet with isotropic equivalent energy $E_{\rm iso}=10^{52}\rm\,erg$, half opening angle $\theta_{\rm j}=0.1\,$rad, initial four-velocity $u_0=100$, and uniform ambient medium density $n=10^{-2}\rm\,cm^{-3}$. We follow the evolution from ultra-relativistic initial conditions until the jet decelerates to non-relativistic speeds. The time for each snapshot is shown in the legend of the upper panel in units of the lab-frame deceleration time $t_{\rm dec}=97\,$d as defined in eq. (\ref{eq:44}), from $t/t_{\rm dec}=0.03$ to 40. In the bottom panel, the arrows show $\theta_{90}$ (in degrees), the angle within which 90\% of the total kinetic energy is contained. Note that, even when the entire jet has decelerated to non-relativistic speeds $\beta\simeq 0.2$, the structure is still non-spherical with $\theta_{90}\sim 40^{\rm o}$.}
\label{fig:tophat}
\end{figure}

\begin{figure}
  \centering
\includegraphics[width = 0.45\textwidth, keepaspectratio]{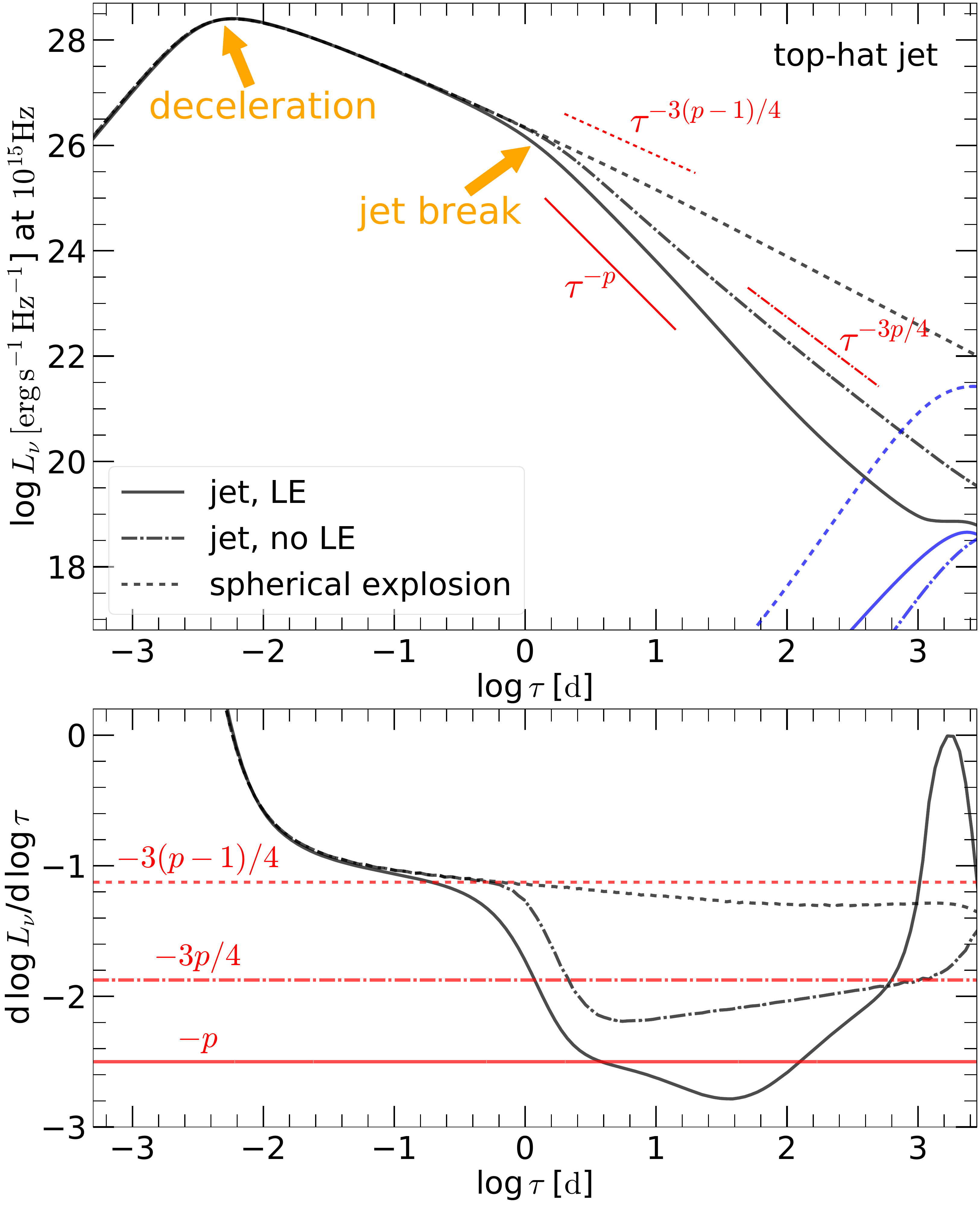} 
\caption{Comparison between afterglow lightcurves from a top-hat jet and spherically symmetric explosion. The line of sight is aligned with the jet axis for all cases and the observer's frequency, $\nu=10^{15}\,$Hz, is chosen to be between $\num$ and $\nu_c$ so as to focus on the jet-break effect. The black lines are for the \textit{total} flux from both forward and counter jets, and the blue lines are for the counter jet (the one moving away from the observer). The solid line is for a top-hat jet of half opening angle $0.1\rm\,rad$ including lateral expansion (LE). The dashed line includes the full emitting sphere of a spherical explosion, and the dash-dotted line only includes the emission from the polar region of a sphere with half opening angle $0.1\rm\,rad$ (no LE). The difference only arises after the jet break at $\tau\sim1\,$d and various analytically expected \citep{1999ApJ...519L..17S} post-jet-break power-law behaviors are marked. The bottom panel shows the numerical slopes, which are known to deviate from analytical power-laws \citep[][and references therein]{2007RMxAC..27..140G}. For all cases, we take isotropic equivalent energy $10^{52}\rm\,erg$, initial four-velocity $u_0=100$, and uniform ambient medium density $n_0=10^{-2}\rm\,cm^{-3}$. Microphysical parameters for the forward shock are $\epse=0.1$, $\epsB=10^{-4}$, $p=2.5$.
}
\label{fig:onaxis}
\end{figure}

\section{Hydrodynamic Results}\label{sec:results}
In this section, we present the results from a few test runs of $\mathtt{Jedi}$. The detailed procedure for synchrotron emission calculation will be presented in the next section.

\begin{figure*}
  \centering
\includegraphics[width = 0.9\textwidth, keepaspectratio]{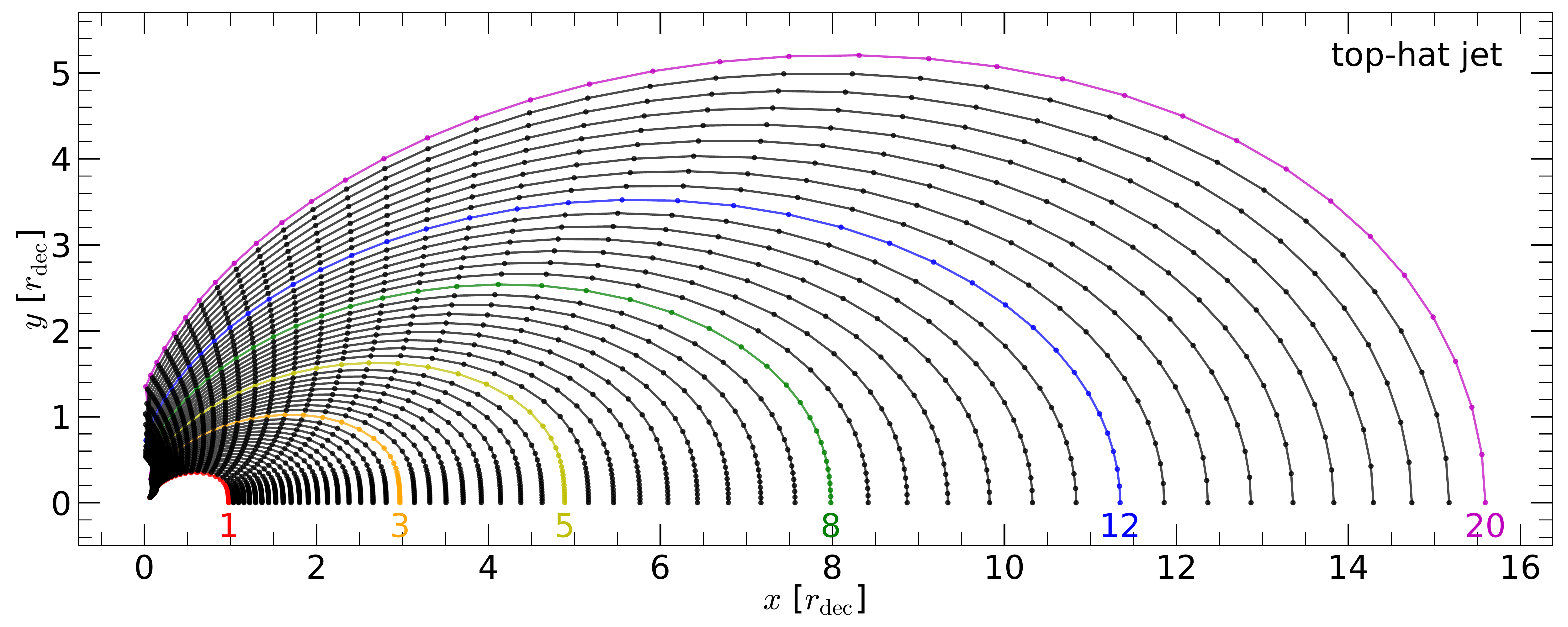} 
\caption{Positions of the jet surface at different epochs from $t/t_{\rm dec}=1$ (innermost red curve) to 20 (outermost magenta curve), for the top-hat jet case described in Fig. \ref{fig:tophat}. Grid points are shown as dots. The jet is axisymmetric with its axis along $\mathbf{\hat{x}}$. The physical units are $r_{\rm dec}=0.081\,$pc and $t_{\rm dec}=97\,$d. 
}\label{fig:tophat_2D}
\end{figure*}

In the current code implementation, we adopt power-law functions for the angular structures of the energy and four-velocity as follows \citep[as considered by][]{1998ApJ...499..301M, 2002MNRAS.332..945R, 2002ApJ...571..876Z, 2003ApJ...591.1075K}
\begin{equation}
    {\d E \over \d \Omega}(\theta) = {E_{\rm iso}\over 4\pi} \left[1 + \left({\theta/ \theta_{\rm c}}\right)^{k} \right]^{-q/k},
\end{equation}
\begin{equation}
    u_0(\theta) = u_{\rm 0,max} \left[1 + \left({\theta/ \theta_{\rm c}}\right)^{k} \right]^{-s/k}.
\end{equation}
We typically choose $k=2$ but larger $k$ can be used for a sharper transition between the jet core and power-law wing.

\subsection{Top-hat jet}

We first show a test case of a (nearly) top-hat jet of angular size $\theta_{\rm j}=0.1\rm\,rad$ with initial conditions given by $E_{\rm iso}=10^{52}\rm\, erg$ and $u_{0, \rm max}=100$. To maintain continuity, we use a sigmoid function to smoothly cutoff the energy near the edge
\begin{equation}
    S(x) = {1-\xi \over 1+\mr{e}^x} + \xi,\ \ x = 50(\theta-\theta_{\rm j}),\ \xi = 10^{-5},
\end{equation}
with the asymptotic behaviors $S\rightarrow 1$  ($S\rightarrow\xi$) as $x\rightarrow -\infty$ ($x\rightarrow+\infty$). Other parameters that are not important for this case are $\theta_{\rm c}=10^{-0.9}\,$rad, $q=s=4$, and $k=5$.

\begin{figure}
  \centering
\includegraphics[width = 0.45\textwidth, keepaspectratio]{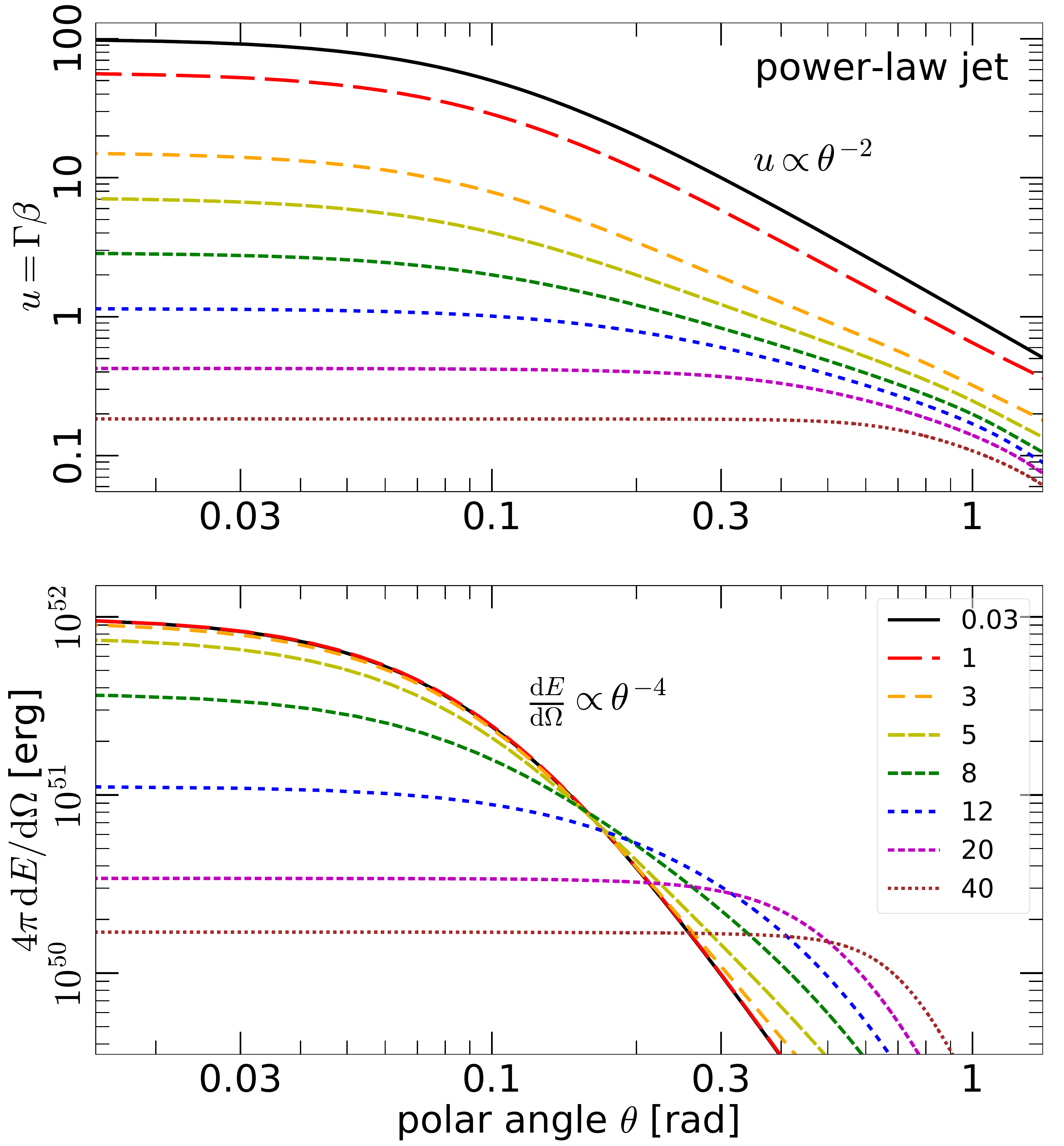}
\caption{Evolution of angular structures $u(\theta)$ (upper panel) and $\d E/\d \Omega(\theta)$ (lower panel) for a power-law jet with peak isotropic energy $E_{\rm iso}=10^{52}\rm\,erg$, core size $\theta_{\rm c}=0.1\,$rad, peak four-velocity $u_{0,\rm max}=100$, energy structure index $q=4$, four-velocity structure index $s=2$, and uniform ambient medium density $n=10^{-2}\rm\,cm^{-3}$. The time for each snapshot is shown in the legend of the lower panel in units of the lab-frame deceleration time $t_{\rm dec}=97\,$d as defined in eq. (\ref{eq:44}), from $t/t_{\rm dec}=0.03$ to 40.
}
\label{fig:PLjet}
\end{figure}

\begin{figure*}
  \centering
\includegraphics[width = 0.9\textwidth, keepaspectratio]{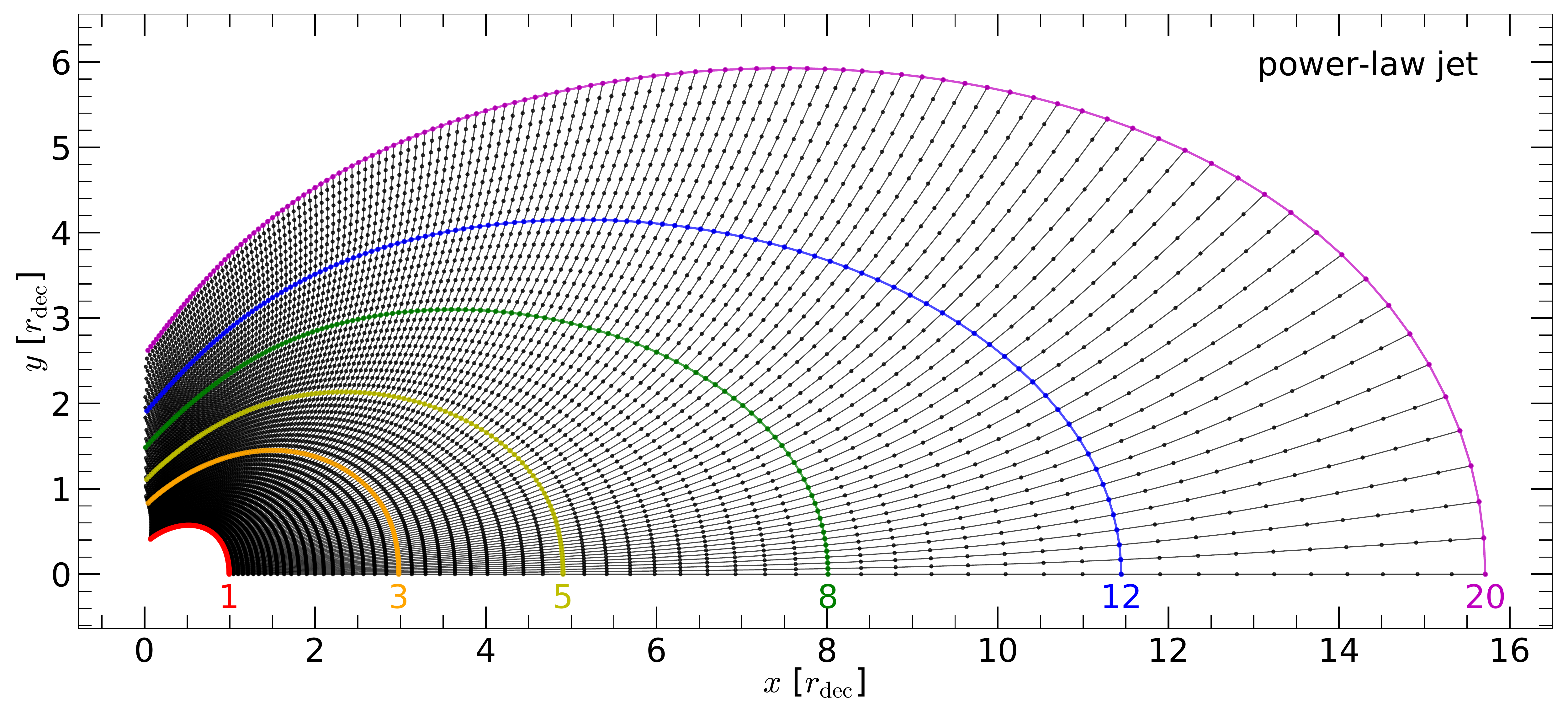}
\caption{Trajectories of the grid points (instead of locations of the jet surface as in Fig. \ref{fig:tophat_2D}) from $t/t_{\rm dec}=1$ to 20 (black curves) for a power-law jet as described in Fig. \ref{fig:PLjet}. We highlight the positions of the jet surface at a number of epochs from $t/t_{\rm dec}=1$ (innermost red curve), 3, 5, 8, 12 to 20 (outermost magenta curve). The physical units are $r_{\rm dec}=0.081\,$pc and $t_{\rm dec}=97\,$d.
}\label{fig:PLjet_2D}
\end{figure*}

\begin{figure*}
  \centering
\includegraphics[width = 0.45\textwidth, keepaspectratio]{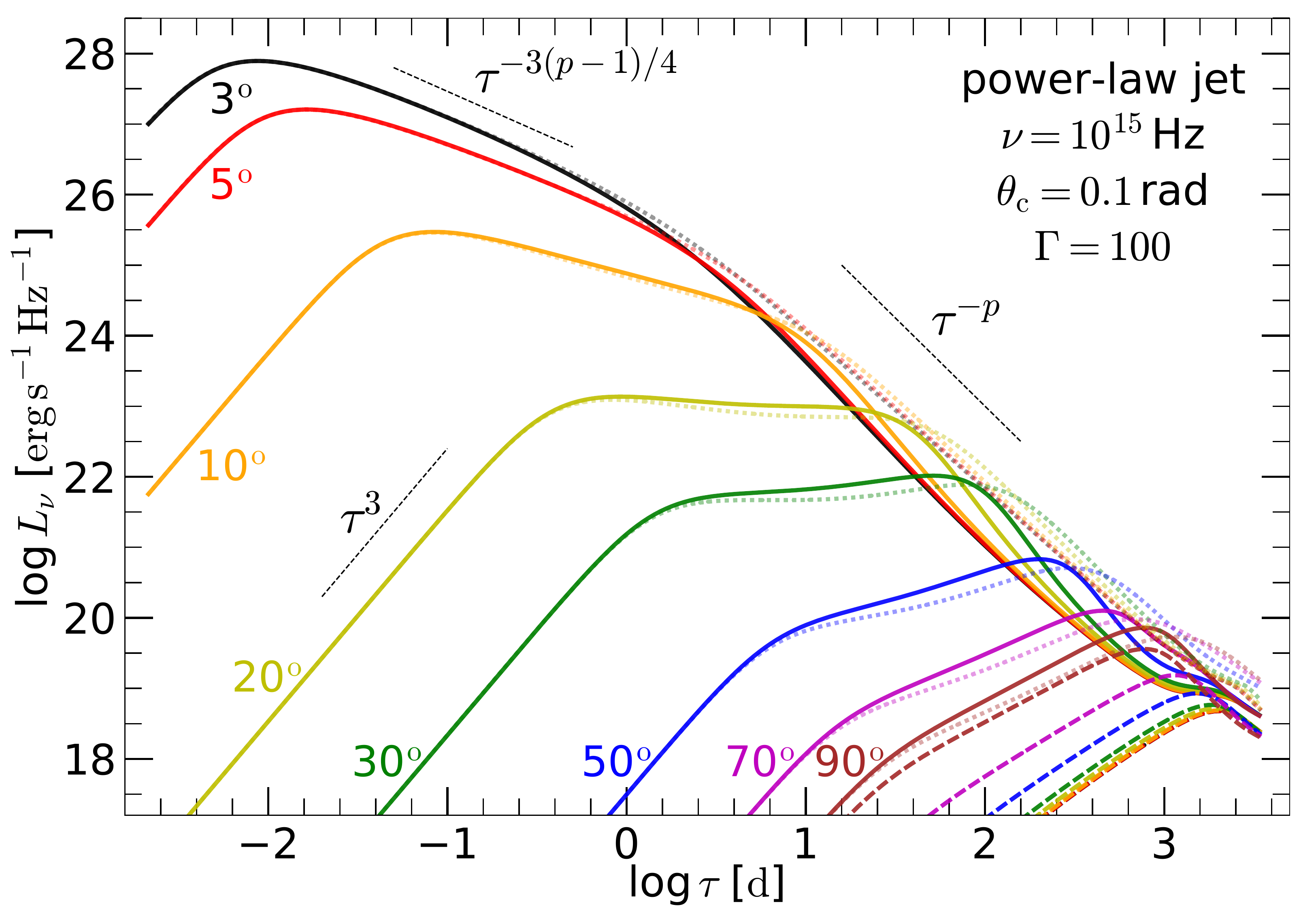} 
\includegraphics[width = 0.45\textwidth, keepaspectratio]{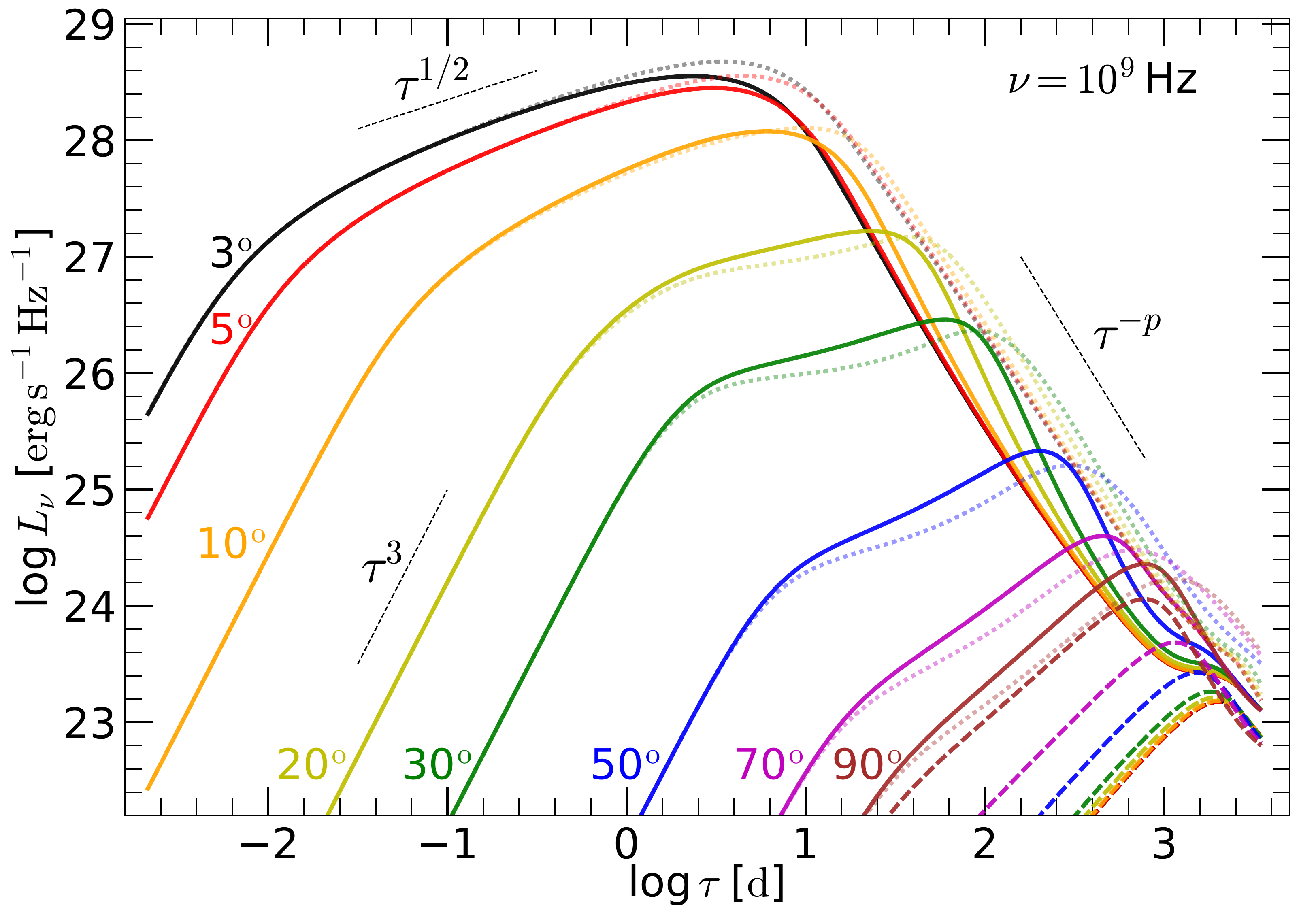} 
\caption{Lightcurves for a power-law jet for different viewing angles from $3^{\rm o}$ (nearly on-axis, black line) to $90^{\rm o}$ (edge-on, brown line) at observer's frequency $\nu=10^{15}\,$Hz (left panel, $\num<\nu<\nuc$) and $\nu=10^9\,$Hz (right panel, $\nu<\num$ in the $L_\nu\propto \tau^{1/2}$ phase and then $\num<\nu<\nuc$ at later time). The solid lines are for \textit{total} flux (including both forward and counter jets), the dashed lines are for contribution from the counter jet only. For the $\theLOS=90^{\rm o}$ case, the total flux is precisely twice of that from each jet. The faint dotted lines are for the total flux from the same jet but without lateral expansion. The initial conditions are the same as in Fig. \ref{fig:PLjet}. Microphysical parameters for the forward shock are $\epse=0.1$, $\epsB=10^{-4}$, $p=2.5$.
}
\label{fig:PLjet_LC}
\end{figure*}

The dynamical evolution for the top-hat jet angular structure at different times are shown in Fig. \ref{fig:tophat}. We show a sequence of the angular structures $\d E/\d \Omega(\theta)$ and $u(\theta)$ at $t/t_{\rm dec}=0.03$ (almost identical to the initial condition), 1 (the jet starts to decelerate), 3, 5 ($\Gamma\sim \theta_{\rm j}^{-1}$, jet break), 8, 12 (trans-relativistic), 20, 40 (non-relativistic). The full simulation runs from lab-frame time $t=0.3\,$d to 6000 d (and the multi-band synchrotron emission is computed) in a few seconds on a single CPU core.

We find that the energy in the jet-core region at $\theta< \theta_{\rm j}$ is only substantially reduced when the Lorentz factor has dropped below $\Gamma\sim 4$ for this case \citep[in agreement with][]{2003ApJ...591.1075K}. Thus, we expect that, for observers near the jet axis (with the model parameters as described above), simplistic afterglow emission model assuming no lateral expansion should be fairly accurate before observer's time $\tau \simeq 8t_{\rm dec}/2\Gamma^2\sim 20\,$d. According to our simulation, the flux under lateral expansion is smaller than that from spherical evolution by a factor of 3 at $\tau=20\,$d, and the discrepancy increases to about an order of magnitude at later time. This is shown in Fig. \ref{fig:onaxis}, where we compare the on-axis synchrotron lightcurves with and without lateral expansion at frequency $\nu=10^{15}\,$Hz (rest-frame UV), which is between $\num$ and $\nuc$ (see \S\ref{sec:synchrotron} for definitions). Microphysical parameters for the forward shock in this and other figures in the paper are taken to be $\epse=0.1$, $\epsB=10^{-4}$, $p=2.5$ in accord with inferences from observed GRB afterglows (e.g. \citealt{2014ApJ...785...29S,2014MNRAS.443.3578N,2015ApJ...806...15Z,2016MNRAS.461...51B,2017MNRAS.472.3161B}). The jet break --- achromatic lightcurve steepening when the edge of (the core of) the jet becomes visible to the observer --- is mainly caused by the \say{missing} contribution to the flux compared to a spherical blastwave, whereas lateral expansion plays a subdominant role. Lateral expansion is more important if the observer's viewing angle is far from the jet axis $\theta_{\rm obs}>\theta_{\rm j}$, e.g., for the observation of \say{orphan} afterglow. A small amount of energy spreading from the core ($<\theta_{\rm j}$) to the wing ($>\theta_{\rm j}$) will brighten the flux at early time. The flux peaks when the core of the jet decelerates to a Lorentz factor $\Gamma\simeq \theta_{\rm obs}^{-1}$, and if $\theta_{\rm obs}>1/4\,$rad (or $15^{\rm o}$), then the peak flux (and peak time) will be substantially affected because the jet energy has already spread out laterally.

Finally, the shape of the jet surface in different snapshots are shown in Fig. \ref{fig:tophat_2D}. We note that even when the jet has decelerated to non-relativistic speeds, the overall shape is still not spherical, in agreement with previous numerical simulations \citep[e.g.,][]{2012ApJ...751...57D}.


\begin{figure*}
  \centering
\includegraphics[width = 0.9\textwidth, keepaspectratio]{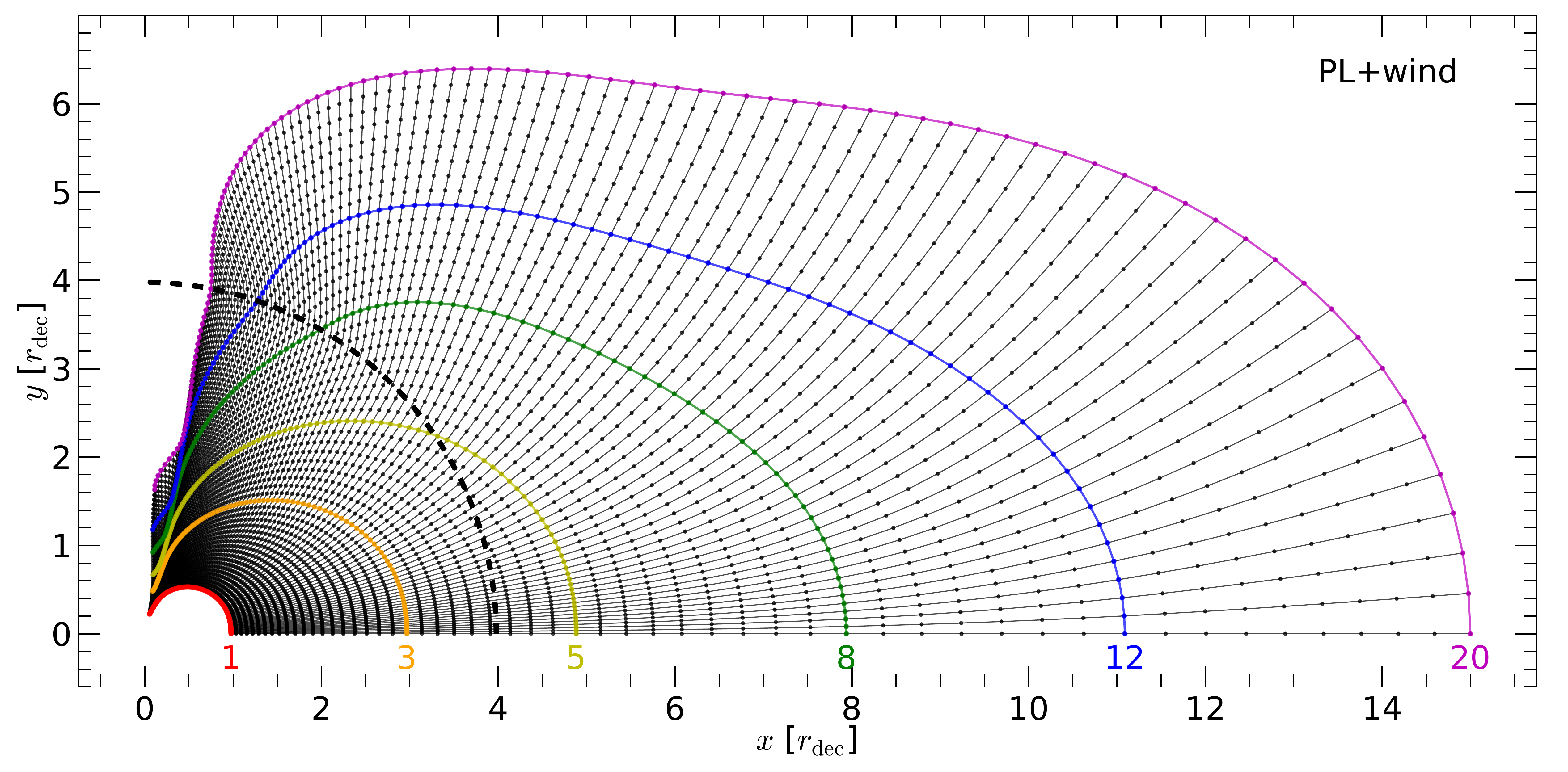}
\caption{Trajectories of the grid points from $t/t_{\rm dec}=1$ to 20 (black curves) for a power-law jet (same as in Fig. \ref{fig:PLjet}) propagating in a wind. We highlight the positions of the jet surface at a number of epochs from $t/t_{\rm dec}=1$ (innermost red curve), 3, 5, 8, 12 to 20 (outermost magenta curve). The physical units are $r_{\rm dec}=0.081\,$pc and $t_{\rm dec}=97\,$d. The wind termination shock at $r_{\rm w}=10^{18}\rm \,cm$ is shown in a thick black dashed line. The density jump (by a factor of 4) at the termination shock causes mild lateral deflections at large angles $\theta\gtrsim 45^{\rm o}$. This causes the late-time shape to be slightly closer to sphericity than the uniform CSM case.
}\label{fig:wind_2D}
\end{figure*}

\subsection{Power-law jet}

We consider another test case of power-law jet structure, which is more realistic than the top-hat case. The power-law index for the energy structure is $q=4$ and for the four-velocity structure is $s=2$. We take $k=2$ for a smooth core-wing transition, the peak isotropic energy $E_{\rm iso}=10^{52}\,$erg, core size $\theta_{\rm c}=0.1\,$rad, peak four-velocity $u_{0,\rm max}=100$, and a constant ambient medium density $n=10^{-2}\rm\, cm^{-3}$. We show the dynamical evolution of the jet structure in Fig. \ref{fig:PLjet}, the numerical trajectories of the grid points in Fig. \ref{fig:PLjet_2D}, and synchrotron emission seen from different viewing angles in Fig. \ref{fig:PLjet_LC} at two different frequencies $\nu=10^9\,$Hz and $10^{15}\,$Hz.

Similar to the top-hat case, the jet core loses significant amount of energy due to lateral expansion only after the Lorentz factor has dropped below about 5, and even when the jet has decelerated to non-relativistic speeds, the global structure approaches sphericity very gradually. We also find that the on-axis lightcurve is similar to the top-hat jet case, but the jet break occurs in a much smoother way. The off-axis lightcurves show a large diversity from shallow decay (viewing angle $\theLOS\lesssim 20^{\rm o}$) to shallow rise ($\theLOS\gtrsim 20^{\rm o}$). The former has been suggested as a possible explanation for X-ray plateaus observed in the lightcurves of cosmological GRBs \citep{Eichler2006,BDDM2020} and the latter is similar to what was seen in GW170817. The quantitative rise and decay slopes depend on and can be used to constrain the jet structure \citep{2018MNRAS.481.1597G,Ghirlanda2019}.

The results for a narrower jet with $\theta_{\rm c}=0.03\,$rad and $u_{\rm 0,max}=300$ (with other parameters the same) are shown in the Appendix. The nontrivial differences from the $\theta_{\rm c}=0.1\,$rad case are (1) lateral expansion occurs earlier and faster (as predicted by eq. \ref{eq:LE} that $v_\perp'/c\sim (\Gamma \theta)^{-1}$), (2) the off-axis lightcurves at $\theLOS\gtrsim 10^{\rm o}$ have faster rises and sharper peaks \citep[in agreement with][]{2018ApJ...868L..11M}. Quantitatively, by the time the Lorentz factor drops below about 10, the energy contained within the jet core has decreased by a factor of 3 for this case. The lightcurves with and without lateral expansion show large differences up to a factor of 30. We conclude that lateral expansion is more important for narrower jets.

\subsection{Wind termination shock}

The third test case we consider has the same power-law angular structure as in the second ($\thec=0.1\,$rad and $u_{\rm 0,max}=100$) case but a non-uniform CSM density profile as follows \citep{2007MNRAS.380.1744N}
\begin{equation}\label{eq:wind_profile}
    \rho_0(\b{r}) =
\begin{cases}
    \rho_{\rm w} (r/r_{\rm w})^{-2}, & \mbox{ for $r<r_{\rm w}$,}\\
    4\rho_{\rm w}, &\mbox{ for $r\geq r_{\rm w}$,}
\end{cases}
\end{equation}
where $r_{\rm w}$ is the radius of the wind termination shock and $\rho_{\rm w}$ is the wind density at $r_{\rm w}$. This is motivated by (1) long/soft GRBs are known to originate from the death of massive He stars at low metallicity \citep{2006ARA&A..44..507W}, and (2) late-time ($t\gtrsim 10\,$hr, the \say{normal decay} segment) afterglow observations favor a constant CSM density for most GRBs rather than a $r^{-2}$ profile \citep[e.g.,][]{2000ApJ...536..195C, 2002ApJ...571..779P}. For a massive star with mass loss rate of $\dot{M} = 10^{-6} \dot{M}_{-6}\rm\, M_\odot \, yr^{-1}$ and wind speed $v_{\rm w} = 3\times 10^8 v_{\rm w, 8.5}\rm\, cm\,s^{-1}$, if the ambient medium confining the wind bubble has pressure $P_{\rm a}=10^{-9}P_{\rm a, -9} \rm\, dyne \, cm^{-2}$, then the radius of the termination shock is given by the pressure balance $\dot{M}v_{\rm w}/(4\pi r_{\rm w}^2) = P_{\rm a}$, which means $r_{\rm w} = 1.2\times10^{18}\mr{\,cm}\, (\dot{M}_{-6}v_{\rm w,8.5}/P_{\rm a, -9})^{1/2}$ \citep{2005ApJ...631..435R, 2006A&A...460..105V}. The large fiducial pressure is expected if the progenitor star is embedded in a cluster of massive stars. For instance, a modest number $N\sim 10^2$ of such windy stars concentrated in a radius of a few parsecs can provide a high-pressure ($\sim 10^{-9} \rm\, dyne \, cm^{-2}$) intra-cluster medium of shocked wind.

\begin{figure}
  \centering
\includegraphics[width = 0.45\textwidth, keepaspectratio]{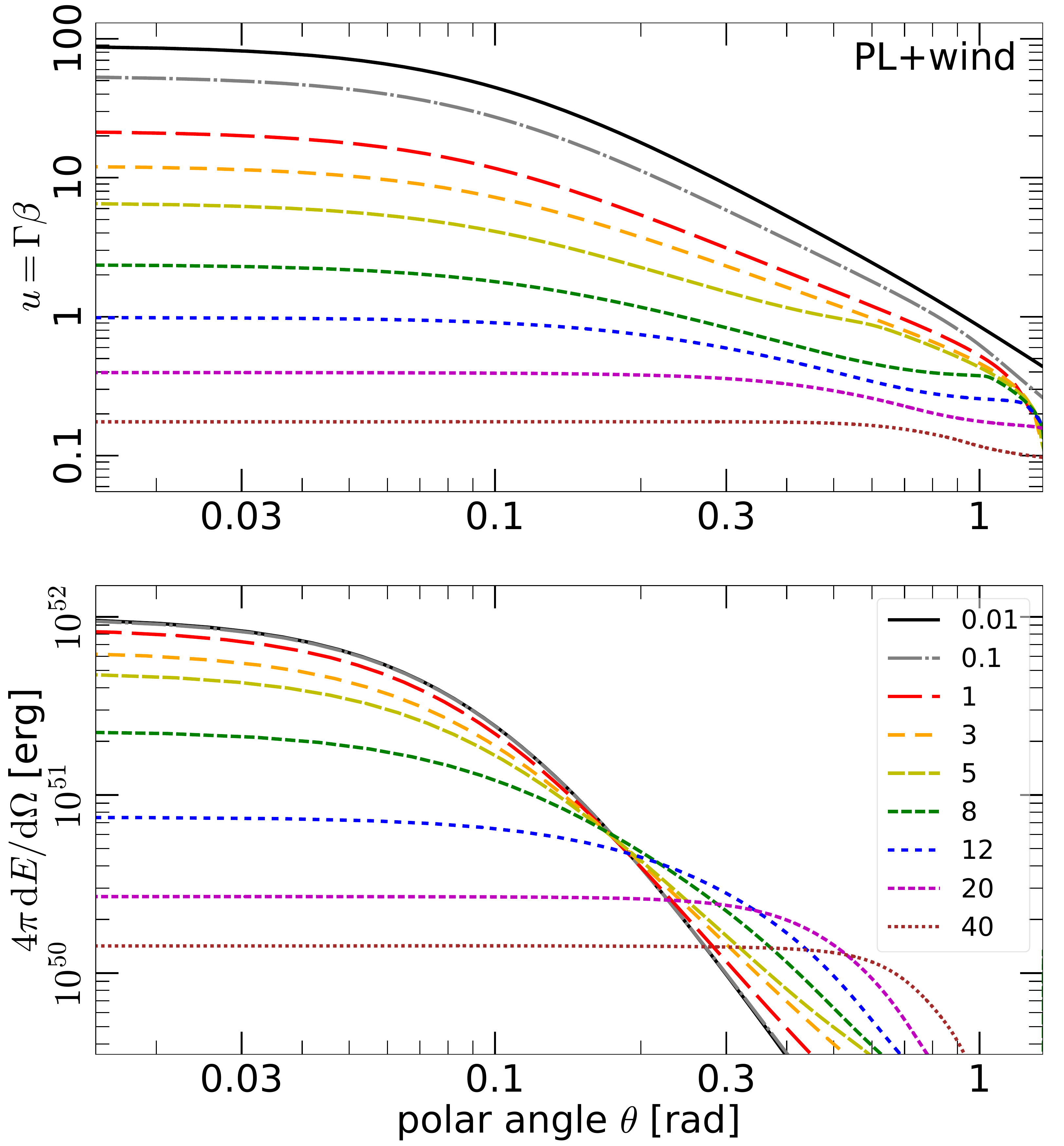}
\caption{Evolution of angular structures $u(\theta)$ (upper panel) and $\d E/\d \Omega(\theta)$ (lower panel) for a power-law jet with peak isotropic energy $E_{\rm iso}=10^{52}\rm\,erg$, core size $\theta_{\rm c}=0.1\,$rad, peak four-velocity $u_{0,\rm max}=100$, energy structure index $q=4$, four-velocity structure index $s=2$. The CSM density profile is a wind with termination shock at $\rw = 10^{18}\,$cm and then uniform ambient medium density $n=10^{-2}\rm\,cm^{-3}$ at larger radii $r>\rw$. The time for each snapshot is shown in the legend of the lower panel in units of the lab-frame deceleration time $t_{\rm dec}=97\,$d as defined in eq. (\ref{eq:44}), from $t/t_{\rm dec}=0.01$ to 40.
}
\label{fig:wind}
\end{figure}

\begin{figure}
  \centering
\includegraphics[width = 0.45\textwidth, keepaspectratio]{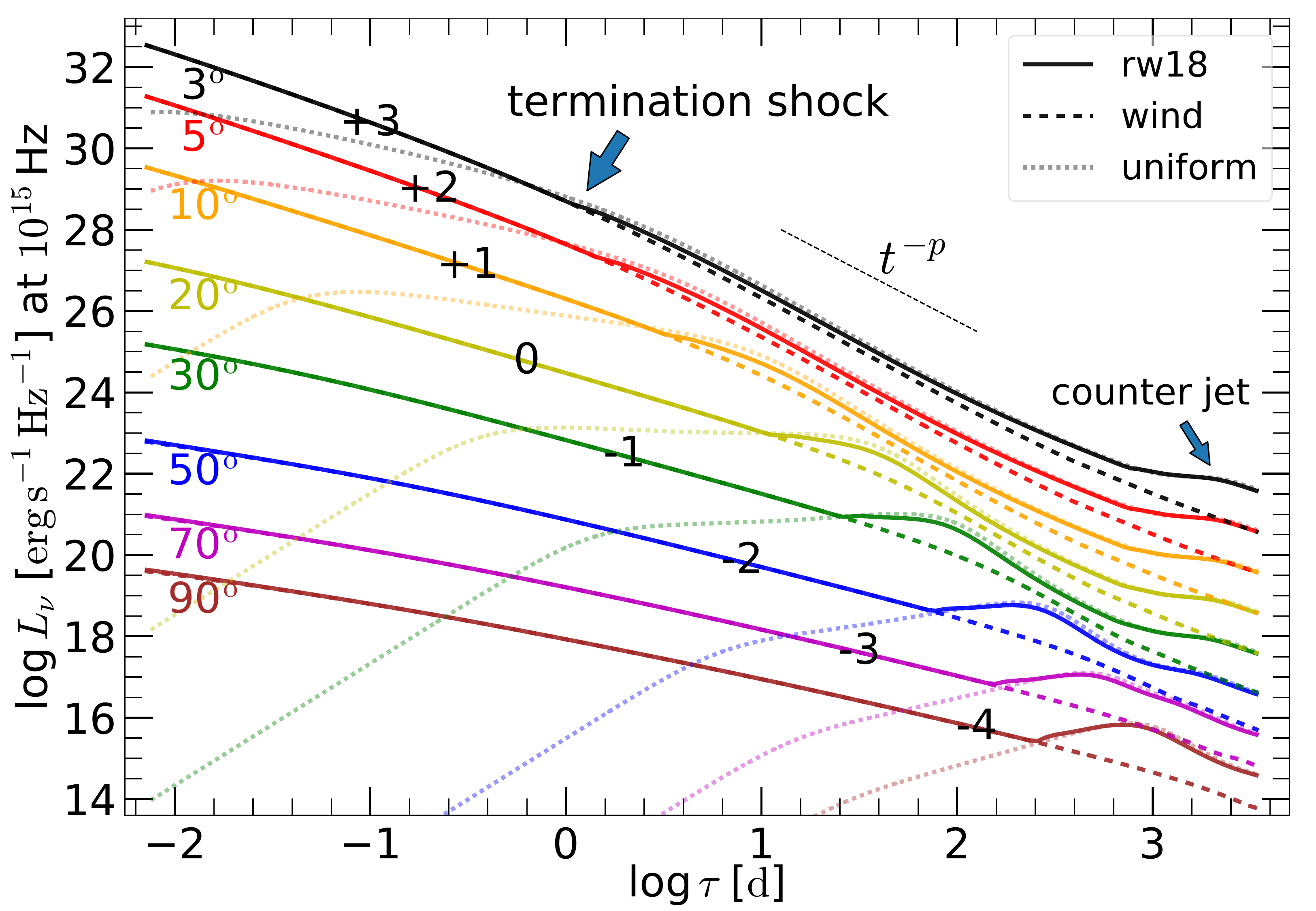} 
\caption{Lightcurves for different viewing angles from $3^{\rm o}$ (nearly on-axis, black line) to $90^{\rm o}$ (edge-on, brown line), with appropriate vertical offsets added for clarity. We focus on the rw18 case with wind termination shock at $\rw=10^{18}\,$cm as shown in solid lines. The dashed lines are for the same wind but without termination shock, and faint dotted lines are for uniform CSM with the same density as that at $r>\rw$ (already shown in Fig. \ref{fig:PLjet_LC}). All three cases have the same jet angular structure (as shown in the left panel of Fig. \ref{fig:PLjet}). Microphysical parameters for the forward shock are $\epse=0.1$, $\epsB=10^{-4}$, $p=2.5$. We see two asymptotic behaviors: at early time before the emitting material reaches the termination shock, the lightcurves are shaped by jet-wind interaction; and at late time after the emitting material has long passed the termination shock, the observer sees similar afterglow emission as in the uniform CSM case.
}
\label{fig:LC_wind}
\end{figure}

Thus, we take $\rw = 10^{18}\rm\, cm$ and $\rho_{\rm w}$ corresponding to number density of $n_{\rm w}=2.5\times 10^{-3}\rm\, cm^{-3}$ in eq. (\ref{eq:wind_profile}) so that the density is the same as the second test case at radius $r>10^{18}\rm\,$cm (for the purpose of better comparison). To avoid numerical noise injection, the density jump at the termination shock is smoothed over $\Delta \mr{log}\,r = 0.01$ using a sigmoid function. The results are compared against (1) a pure wind CSM case without termination shock and (2) the uniform CSM case as considered earlier.

A caveat for simulating a power-law structured jet interacting with a wind CSM profile is that deceleration at large polar angles $\theta\gg \theta_{\rm c}$ may be much faster than near the jet axis. In the absence of lateral expansion, the deceleration radius scales as $\rdec\propto (\d E/\d \Omega) u_0^{-2}\propto \theta^{2s-q}$ for a wind profile $n\propto r^{-2}$ as compared to $\rdec\propto \theta^{(2s-q)/3}$ for the constant density case. In reality, the materials far from the jet axis have lower Lorentz factors, undergo rapid lateral expansion, and hence decelerate rapidly. To capture this rapid deceleration, which is important for the lightcurves at large viewing angles $\theLOS\gtrsim 50^{\rm o}$, the simulations must start at time $t_0\lll t_{\rm dec}$. Here $t_{\rm dec}$ is our machine time unit as given by eq. (\ref{eq:44}) and we take the density normalization to be $\rho_{\rm norm}=4\rho_{\rm w}$. In practice, we obtained good convergence by taking $t_0/t_{\rm dec} = 10^{-5}$ or smaller, and since our time stepping is logarithmic, this only slightly increases the computational cost.

The results for this case are shown in Figs. \ref{fig:wind_2D}, \ref{fig:wind}, \ref{fig:LC_wind}. Compared to the constant density case, deceleration and lateral expansion occur earlier (before reaching the termination shock). By the time the jet Lorentz factor decreases to 10, the energy contained within the jet core has dropped by a factor of 2. At very late time $t\gtrsim 20 t_{\rm dec}$, most parts of the jet have passed the termination shock and the angular structure is slightly more spherical than but overall quite similar to that of the uniform CSM case. The lightcurves are initially shaped by jet-wind interaction and the fluxes at all viewing angles are much brighter than the uniform CSM case before the emitting material hits the termination shock. The density jump (by a factor of 4) at the termination shock generally brightens the afterglow emission, but the flux enhancement is very mild and smooth for small viewing angles $\theLOS\lesssim 10^{\rm o}$ as is typically the case for most GRBs discovered by prompt $\gamma$-ray emission. This is in agreement with the results of \citet{2007MNRAS.380.1744N}, see also \citet{2007ApJ...665L..93U, 2014ApJ...789...39U}. However, at larger viewing angles $\theLOS\gtrsim 30^{\rm o}$, the flux enhancement due to termination shock is noticeable in the form of a smooth bump. The height of the bump is roughly given by the flux peak seen from far off-axis in the uniform CSM case. This signature should be searched for in future nearby off-axis GRBs.

\subsection{Boosted fireball and comparison to $\mathtt{JetFit}$}

\begin{figure}
  \centering
\includegraphics[width = 0.45\textwidth, keepaspectratio]{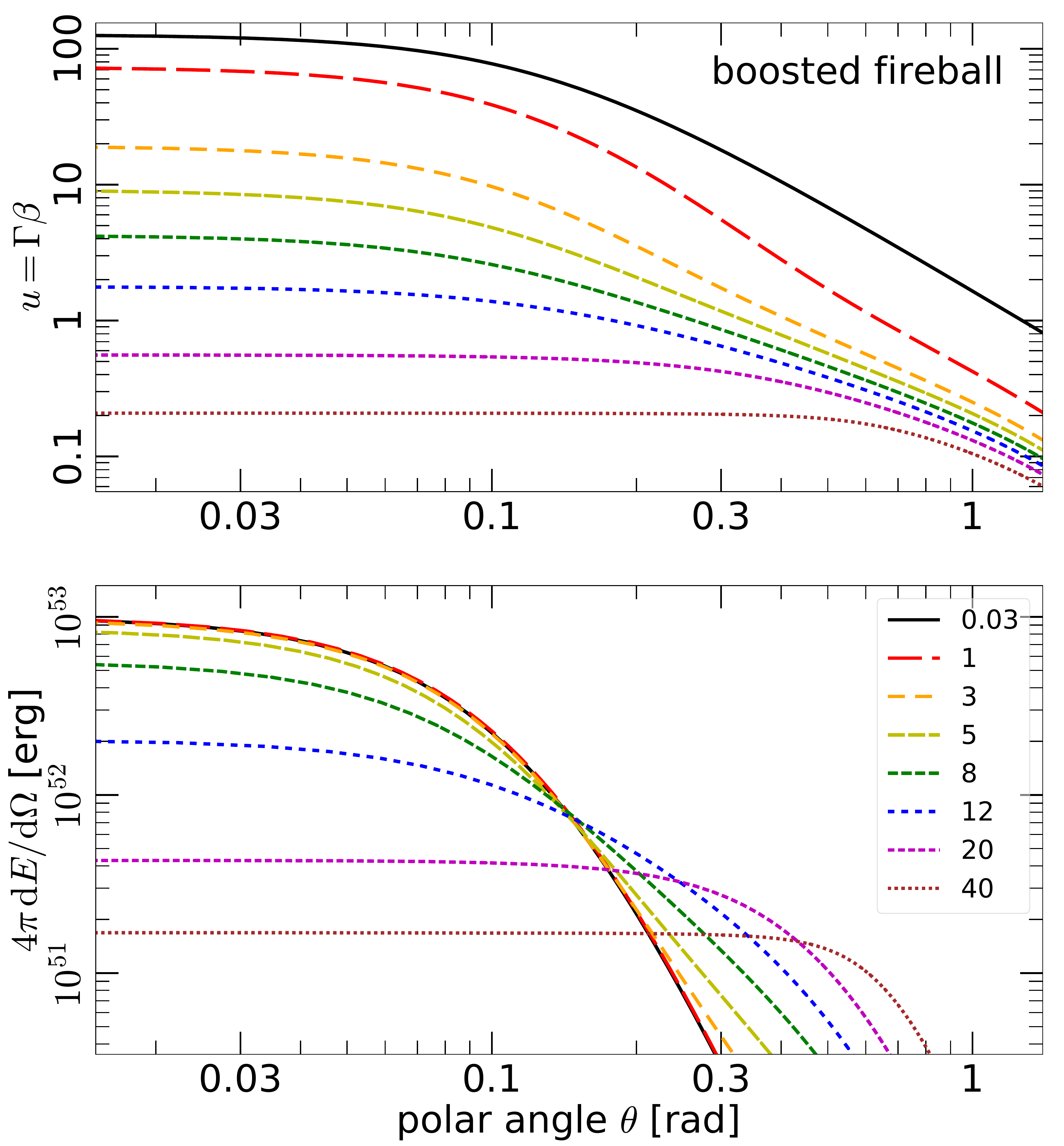}
\caption{Evolution of angular structures $u(\theta)$ (upper panel) and $\d E/\d \Omega(\theta)$ (lower panel) for a boosted-fireball model of \citet{2013ApJ...776L...9D} with $\eta_0=\gamma_{\rm B}=8$, peak isotropic energy $E_{\rm iso}=10^{53}\rm\,erg$, and uniform ambient medium density $n=10^{-2}\rm\,cm^{-3}$. The time for each snapshot is shown in the legend of the lower panel in units of the lab-frame deceleration time $t_{\rm dec}=178\,$d as defined in eq. (\ref{eq:44}), from $t/t_{\rm dec}=0.03$ to 40.
}
\label{fig:bfb_structure}
\end{figure}

\begin{figure}
  \centering
\includegraphics[width = 0.45\textwidth, keepaspectratio]{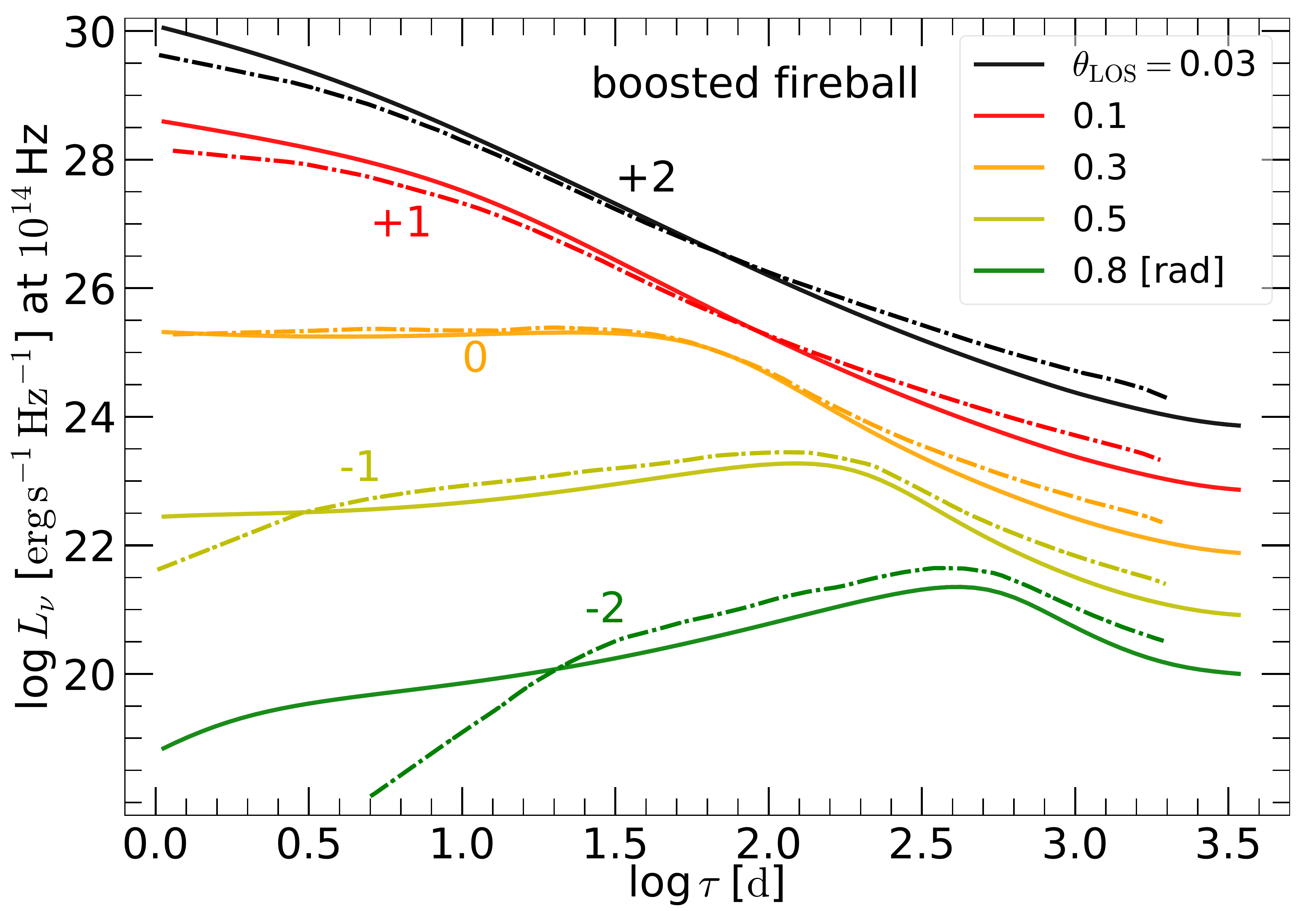} 
\caption{Lightcurves for a boosted-fireball jet for different viewing angles from $\theLOS=0.03$ (black) to $0.8\,$rad (green lines), including contributions from both forward and counter jets. The dash-dotted lines show the lightcurves of the corresponding cases from 2D hydrodynamic simulation \citep{2018ApJ...869...55W}. Microphysical parameters for the forward shock are $\epse=0.1$, $\epsB=10^{-4}$, $p=2.15$.
}
\label{fig:LC_bfb}
\end{figure}

The last test case we consider is the \say{boosted fireball} model used by \citet{2013ApJ...776L...9D, 2018ApJ...869...55W}. This is for the purpose of comparing our numerical results to that from more realistic 2D simulations. The \say{boosted fireball} structure is described by 2 parameters $\eta_0$ (the Lorentz factor in the center-of-mass comoving frame) and $\gamma_{\rm B}$ (the linear boost), which give peak Lorentz factor $2\eta_0\gamma_{\rm B}$ and jet opening angle $\sim \gamma_{\rm B}^{-1}$ \citep{2013ApJ...776L...9D}. Our fiducial power-law model is described by 4 parameters $\theta_{\rm c}$, $q$, $s$ and $u_{\rm 0, max}$ and is hence more flexible\footnote{We note that not all \say{boosted fireball} structures can be well described by a power-law model. It is unclear which choice is more physical, because the realistic jet structure depends on both the central engine properties and the interaction with a cocoon, which is due to the jet punching through the star or dense material in the immediate vicinity of the compact object \citep[e.g.,][]{2002MNRAS.337.1349R, 2009ApJ...700L..47L, 2011ApJ...740..100B, 2018MNRAS.478.4553D, 2018MNRAS.479..588G, 2018MNRAS.473L.121K}.}.

Our test case has $\eta_0=\gamma_{\rm B}=8$, and the corresponding best-match power-law model parameters are $\theta_{\rm c}=0.13\,$rad, $q = 6.24$, $s=2.12$, $k=2$, and $u_{0,\rm max}=127$. The evolution of the angular structure is shown in Fig. \ref{fig:bfb_structure}. Then in Fig. \ref{fig:LC_bfb}, we compare the lightcurves from $\mathtt{Jedi}$ with the corresponding cases as computed by $\mathtt{JetFit}$ \citep{2018ApJ...869...55W}, which is based on 2D numerical simulations. We find reasonable agreement within a factor of 2 for small viewing angles $\theLOS\lesssim 0.3\,$rad, but for larger $\theLOS$, there is a discrepancy in the early flux rise. This is likely because when $\mathtt{JetFit}$ compresses the 2D numerical simulation results into the so-called \say{Box} snapshots (to reduce memory demand), there is a maximum polar angle $\theta_{99}$ within which 99\% of the total energy is contained \citep{2012ApJ...749...44V}. At early time when lateral expansion is not significant, we have $\theta_{99}\simeq 0.35\,$rad for the current case considered. Physically, the early time flux comes from the far off-axis part of the jet at $\theta\sim \theLOS$ (although this region may only contain a small fraction of the total jet energy), and at later time as each part of the jet decelerates, the observer sees flux contributions from smaller and smaller polar angles \citep[e.g.,][]{2018MNRAS.481.1597G, 2019arXiv190911691R, 2020MNRAS.493.3521B}. This means that the early time emission from larger viewing angles $\theLOS\gtrsim \theta_{99}$ may be strongly affected by the \say{Box} compression. Near the peak time when most of the flux comes from near the jet core, the results between $\mathtt{Jedi}$ and $\mathtt{JetFit}$ agree within a factor of about 2. This small difference may be due to the fact that the jet energy is not concentrated in an infinitely thin shell in the \say{boosted fireball} structure \citep[see Fig. 1 of][]{2018ApJ...869...55W} and that these two codes compute synchrotron emission in slightly different ways.

To conclude this section, we find that our simplified and highly efficient method provides a good approximation for the hydrodynamics of relativistic jets and their synchrotron emission for a wide variety of jet structures and CSM density profiles. 

\section{Synchrotron Emission}\label{sec:synchrotron}
In this section, we calculate the synchrotron emission from the CSM swept up and heated by the forward shock, based on the standard afterglow theory \citep[see][for a recent review]{2015PhR...561....1K}. We provide a full description of our method for completeness. The qualitative improvements upon previous works based on hydrodynamic simulations are: (1) synchrotron self-absorption is taken into account self-consistently, and (2) the spectral shape near break frequencies ($\nu_{\rm a}$, $\num$, and $\nuc$) are calculated in an accurate way (based on the assumption of broken power-law electron Lorentz factor distribution).

We describe the procedure for a particular Lagrangian fluid element, with its mass gaining history $M(t)$, trajectory $\b{r}(t)$, $\b{u}(t)$, and surface area evolution $A(t)$. Any quantity $Q$ in the fluid comoving frame is denoted as $Q'$. The total flux is the sum of all fluid elements. The observer's line of sight (LOS) is placed at an angle $\theLOS$ with respect to the jet axis. We consider the observer to be located in the host galaxy rest frame (it is straightforward to include a cosmological redshift). At observer's time $\tau$, we (linearly) interpolate the numerical trajectory of the fluid element $\b{r}(t)$ to find the position and lab-frame time $t$ corresponding to the photon arrival time,
\begin{equation}
  \label{eq:3}
\tau = t - (r/c)\cos\theobs,
\end{equation}
where $\theobs$ is the angle between the LOS and the radial vector $\b{r}$, given by
\begin{equation}
  \label{eq:20}
  \cos\theobs = \sin \theLOS \sin \theta \cos \phi + \cos \theLOS \cos
  \theta,
\end{equation}
where $\theta$ is the polar angle between $\b{r}$ and the jet axis and $\phi$ is the azimuthal angle. The shocked region for each fluid element is considered as a slab whose normal direction is parallel to its velocity vector $\b{u}(t)$. The specific intensity at observer's frequency $\nu$ on the surface of the slab in the observer's direction is denoted as $I_\nu$, which is directly related to the observed flux. The angle between the LOS and $\b{u}$ is denoted as $\btheobs$ in the lab frame and $\btheobs'$ in the comoving frame, given by
\begin{equation}
  \label{eq:38}
  \cos\btheobs = \sin \theLOS \sin \theta_u \cos \phi + \cos \theLOS \cos
  \theta_u,
\end{equation}
and
\begin{equation}
    \cos \btheobs' = {\beta - \cos\btheobs \over 1 - \beta \cos\btheobs},
\end{equation}
where $\theta_u$ is the angle between the velocity vector and the jet axis. Note that $\theta_u$ is generally different from $\theta$ due to lateral expansion. Thus, the Doppler factor is given by
\begin{equation}
  \label{eq:21}
  \mc{D} = \left[\Gamma(1 - \beta \cos\btheobs)\right]^{-1}.
\end{equation}
The specific intensity in the lab frame is related to that in the comoving frame $I_{\nu'}'$ by
\begin{equation}
\label{eq:intensity_lab}
    I_\nu = I'_{\nu'} \mc{D}^3,\ \ \nu = \nu' \mc{D}.
\end{equation}
Under the assumption of a uniform slab, the intensity in the comoving frame is given by the solution of one-dimensional radiative transfer \citep[][eq. 1.30]{1979rpa..book.....R}
\begin{equation}
\label{eq:intensity_comoving}
    I'_{\nu'} = (1 - \mr{e}^{-\tau_{\nu'}}) {\bar{P}_{\nu'} \over 4\pi \bar{\sigma}_{\nu'}},
\end{equation}
where $\bar{P}_{\nu'}$ is the average specific power per electron in the fluid comoving frame (emissivity $=P_{\nu'}/4\pi$ assuming isotropic plasma), $\bar{\sigma}_{\nu'}$ is the averaged synchrotron self-absorption cross section, and $\tau_{\nu'}$ is the optical depth along the direction of the LOS (not to be confused with the observer's time in earlier sections), given by
\begin{equation}
    \tau_{\nu'} = {\bar{\sigma}_{\nu'} N_e\over |\cos\btheobs'| A},
\end{equation}
and $N_e$ is the total number of relativistic electrons. Then, the isotropic equivalent specific luminosity at frequency $\nu$ contributed by this fluid element is given by
(see proof in the Appendix)
\begin{equation}
\label{eq:luminosity}
    L_\nu = 4\pi I_\nu A |\cos\btheobs'|.
\end{equation}
Given the host galaxy's redshift $z_{\rm host}$ and luminosity distance $D_{\rm L}$, the specific luminosity $L_\nu$ can be easily converted to flux density $F_{\nu_{\rm obs}}$ at the redshifted frequency $\nu_{\rm obs}=\nu/(1+z_{\rm host})$ by $\nu_{\rm obs} F_{\nu_{\rm obs}} = \nu L_\nu/(4\pi D_{\rm L}^2)$.


In the following, we discuss the electron Lorentz factor distribution and magnetic fields of the shocked CSM, following the standard procedure \citep[e.g.][]{1998ApJ...497L..17S}.

We assume that electrons and magnetic fields share fractions $\epse$ and $\epsB$ of the thermal energy density of the shocked CSM ($e'-\rho'c^2$, not including rest-mass).  The magnetic field strength in the comoving frame is
\begin{equation}
  \label{eq:19}
  B' = \left[32\pi \Gamma (\Gamma-1) \epsB \rho_0(\b{r}) c^2\right]^{1/2},
\end{equation}
where $\rho_0(\b{r})$ is the density of the pre-shock CSM at position $\b{r}$. Electrons are accelerated to a power-law momentum distribution of index $p$. Since generally $2<p<3$, the majority of kinetic energy is in the lowest energy but relativistic particles, so the minimum Lorentz factor is taken as \citep{2006ApJ...638..391G, 2013ApJ...778..107S}
\begin{equation}
  \label{eq:17}
  \gm = \mr{max}\left[2, \ 
(\Gamma -1) \epse {p-2\over p-1} {\mp \over \me}\right].
\end{equation}
Note that electron Lorentz factor in the fluid comoving frame is denoted as $\gamma$ without a prime (since there is no possible confusion). Radiative cooling is important for electrons above the Lorentz factor $\gc$, which is given by equating the dynamical time $t_{\rm dy}'=t/2\Gamma$ to the synchrotron cooling time $t_{\rm c}' = 6\pi \me c^2/(\gamma \sigmaT c B'^2)$,
\begin{equation}
\label{eq:gamma_c}
    \gc = {12\pi \Gamma \me c \over B'^2 t \sigmaT},
\end{equation}
where $\sigmaT$ is the Thomson scattering cross section. Since the thermal energy density of the (one-zone) shocked region is dictated by the forward-shock jump condition at each moment, we do not self-consistently take into account adiabatic cooling. We are making only a small error because the number of adiabatically cooled electrons (injected before $\sim t/2$) and their synchrotron flux are subdominant compared to the freshly injected ones (from $\sim t/2$ to $t$). The total number of relativistic electrons in a fluid element is given by the total energy of relativistic electrons ($\gamma>\gm$) being $(\Gamma-1)\epse Mc^2$, i.e.
\begin{equation}
\label{eq:Ne}
    N_e = {p-2\over p-1} (\Gamma-1)\epse {M\over \gm \me}.
\end{equation}
Note that if $\gm>2$, then $N_e = M/\mp$, meaning that all electrons are relativistic. The shape of the electron Lorentz factor distribution is complex near the transition at $\gc$, and we use the following simplified broken power-law ($N_\gamma \equiv \d N/\d\gamma$)
\begin{equation}
\label{eq:LF_distribution}
   N_\gamma = 
\begin{cases}
N_e{p-1\over \gm} (\gamma/\gm)^{-p}\ \ \mbox{for $\gm<\gamma<\gc$,}\\
N_e{(p-1)\gc\over \gm^2} (\gamma/\gm)^{-1-p}\ \ \mbox{for $\gamma>\gc$.}
\end{cases}
\end{equation}
The fast cooling case with $\gc < \gm$ is yet to be implemented in a future version of the code. We also note that the current implementation does not take into account inverse-Compton emission and cooling (by scattering synchrotron or external photons). If $(\epse/\epsB) (\gc/\gm)^{2-p}\gtrsim 1$ (since a fraction $(\gc/\gm)^{2-p}$ of the energy shared by electrons is radiated as photons), synchrotron self-Compton (SSC) cooling reduces $\gc$ and modifies the electron Lorentz factor distribution $N_{\gamma}$ above $\gc$. A self-consistent treatment of SSC is needed to correctly model the afterglow at high frequencies \citep{2001ApJ...548..787S, 2009ApJ...703..675N, 2015MNRAS.454.1073B}.

The specific synchrotron power at $\nu'$ for an electron of Lorentz factor $\gamma$, averaged over an isotropic distribution of pitch angles $\alpha$, is given by
\begin{equation}
    P_{\nu'} = {\sqrt{3} e^3 B \over \me c^2} \t{F}(\nu'/\nu_{\rm syn}'),\ \ \nu_{\rm syn}' = {3\gamma^2 e B'\over 4\pi \me c},
\end{equation}
\begin{equation}
    \t{F}(x) = \int_0^{\pi/2} F\left({x\over \sin\alpha}\right) \sin^2\alpha\, \d\alpha,
\end{equation}
where $F(x)=x\int_x^\infty K_{5/3}(z)\d z$ is the synchrotron function described by modified Bessel functions. Using the asymptotic behavior of $F(x)$ given by \citet{1965ARA&A...3..297G}, we obtain
\begin{equation}
    \t{F}(x) \approx
\begin{cases}
a_0x^{1/3} (1 + a_1 x^{2/3} + a_2x^2),\ \ \mbox{for $x\ll 1$,}\\
a_3\,x^{1/2} \mr{e}^{-x},\ \ \mbox{for $x\gg 1$},
\end{cases}
\end{equation}
where $a_0=1.8084$, $a_1 = -1.0030$, $a_2=0.46875$, $a_3=2.8132$.
The values in between $x\in (10^{-3}, 30)$ are computed for a fine numerical grid and then interpolated (in log space) to arbitrary $x$ with the linear method. For the broken power-law Lorentz factor distribution in eq. (\ref{eq:LF_distribution}), the average specific power  per electron  at frequency $\nu'$ is given by
\begin{equation}
\begin{split}
    \bar{P}_{\nu'} = {1\over N_e}\int_{\gm}^\infty \d \gamma \, N_\gamma P_{\nu'} = {\sqrt{3} e^3 B \over \me c^2} I_1.
\end{split}
\end{equation}
The average absorption cross section at frequency $\nu'$ is given by \citep[][eq. 6.50]{1979rpa..book.....R}
\begin{equation}
\small
\begin{split}
    \bar{\sigma}_{\nu'} = {-c^2\over 8\pi N_e \nu'^2} \int_{\gm}^\infty \d \gamma P_{\nu'} \gamma^2 {\d \over \d\gamma} \left(N_\gamma \over \gamma^2\right)
    = {\sqrt{3} e^3 B'\over 8\pi \me \nu'^2 \gm} I_2.
\end{split}
\end{equation}
For convenience, we have defined the following functions
\begin{equation}
\begin{split}
    I_1(\xc, \xm, p) =&\, {p-1\over 2} \xm^{-{p\over 2}+{1\over 2}} (g_3(\xm) - g_3(\xc)) \\
        &+ {p-1\over 2} \left(\xm\over \xc\right)^{1/2} \xm^{-p/2} g_2(\xc),
\end{split}
\end{equation}
\begin{equation}
\begin{split}
    I_2(\xc, \xm, p) =&\, {p-1\over 2(p+2)} \xm^{-p/2} (g_2(\xm)-g_2(\xc)) \\
    &+ {p-1\over 2(p+3)} \left(\xm\over \xc\right)^{1/2} \xm^{-{p\over 2}-{1\over 2}} g_1(\xc),
\end{split}
\end{equation}
\begin{equation}
    \xm = \nu'/\num',\ \xc = \nu'/\nuc',\ \nu_{\rm m/c}' = \gamma_{\rm m/c}^2{3eB'\over 4\pi \me c},
\end{equation}
\begin{equation}
    g_n(p, x) = \int_0^x x^{(p-n)/2} \t{F}(x) \d x,\ \ n=1,\, 2,\, 3.
\end{equation}
The asymptotic behavior of $g_n$ in the limit $x\ll 1$ is
\begin{equation}
\small
    g_{n}(p,x) = 2a_0\left({x^{(3p-3n+8)/6}\over p-n+8/3} + {a_1 x^{(p-n+4)/2} \over p-n+4} + {a_2 x^{(3p-3n+20)/6} \over p-n+20/3} \right). 
\end{equation}
In the other limit $x\gg 1$, $g_n(p, x)\rightarrow g_{n}(p, \infty)$ (independent of $x$). For $x\in (10^{-3}, 30)$, we compute the numerical values of $g_n(p, x)$ (for $n=1, 2, 3$) for a fine 2D grid and then interpolate to arbitrary values of $x$ (in log space) and $p\in(2, 3)$ with the bilinear method. For $x<10^{-3}$ or $x>30$, we use the above asymptotic functions.

Therefore, we can efficiently calculate the numerical values of $\bar{P}_{\nu'}$ and $\bar{\sigma}_{\nu'}$ at any frequency $\nu'$, for any broken power-law electron Lorentz factor distribution whose shape is described by $\gm$, $\gc$, and $p$ according to eq. (\ref{eq:LF_distribution}). Then we obtain the specific intensity on the surface of the fluid element according to eqs. (\ref{eq:intensity_comoving}) and (\ref{eq:intensity_lab}), and finally the luminosity by eq. (\ref{eq:luminosity}). In the Appendix, we also provide a method to calculate the proper motion of the flux centroid, which is useful for sufficiently nearby jets viewed from off-axis.






\section{Summary and Discussion}\label{sec:summary}
We have presented a new model for the hydrodynamics of a relativistic jet interacting with the surrounding medium, taking into account lateral expansion. The 2D axisymmetric hydrodynamic problem is simplified into a 1D calculation by assuming that the jet and swept-up CSM material are confined in an infinitely thin surface. We derive the equation of motion for each fluid element on the surface from basic conservation laws. The pressure of the shocked CSM is given by the jump conditions at the forward shock, and the pressure gradient between neighboring fluid elements drives lateral expansion. The method is implemented in a numerical (C++) code $\mathtt{Jedi}$, which solves the jet evolution from ultra-relativistic initial conditions to non-relativistic speeds, as well as the synchrotron flux at arbitrary viewing angles and frequencies, in a few seconds on a single CPU core. We have demonstrated in a number of test cases that our method provides a good approximation for the hydrodynamics and afterglow emission for a wide variety of jet structures and CSM density profiles.

The majority of GRBs discovered by prompt $\gamma$-ray emission have their jet axis nearly aligned with our line of sight \citep{BN2019}. It is difficult to constrain their angular structures beyond the jet core which contains most of the energy and dominates the afterglow flux at all time. Recently, joint gravitational wave-electromagnetic detections of GW170817 from an off-axis viewing angle \citep{2017ApJ...848L..12A} makes it possible to learn about the jet structure because the flux before $150\,$d is dominated by large polar angle regions far from the jet axis \citep[e.g.,][]{2018PhRvL.120x1103L, 2018MNRAS.481.1597G}. Valuable constraints on the jet structure may also be obtained by the properties of the prompt emission that will be seen in the population of future GW detected short GRBs \citep{BPBG2019}. Our model provide an efficient way to constrain the jet structure and CSM density profile of similar off-axis events. We will present the application to the afterglow of GW170817 in a separate paper.


\section{acknowledgments}
We thank Chris White, Bing Zhang, Andrew MacFadyen, Ore Gottlieb, Kunal Mooley, Cl{\'e}ment Bonnerot, and Saul Teukolsky for useful discussions. 
We acknowledge the Texas Advanced Computing Center (TACC) at The University of Texas at Austin for providing HPC resources that have contributed to the research results reported within this paper. This research benefited from interactions at the ZTF Theory Network Meeting, funded by the National Science Foundation under Grant No. NSF PHY-1748958. WL was supported by the David and Ellen Lee Fellowship at Caltech. The research of PB was funded by the Gordon and Betty Moore Foundation through Grant GBMF5076.

{\small
\bibliographystyle{mnras}
\bibliography{refs}
}

\appendix

\section{Apparent area of an inclined surface at relativistic speed}
\begin{figure}
  \centering
\includegraphics[width = 0.35\textwidth]{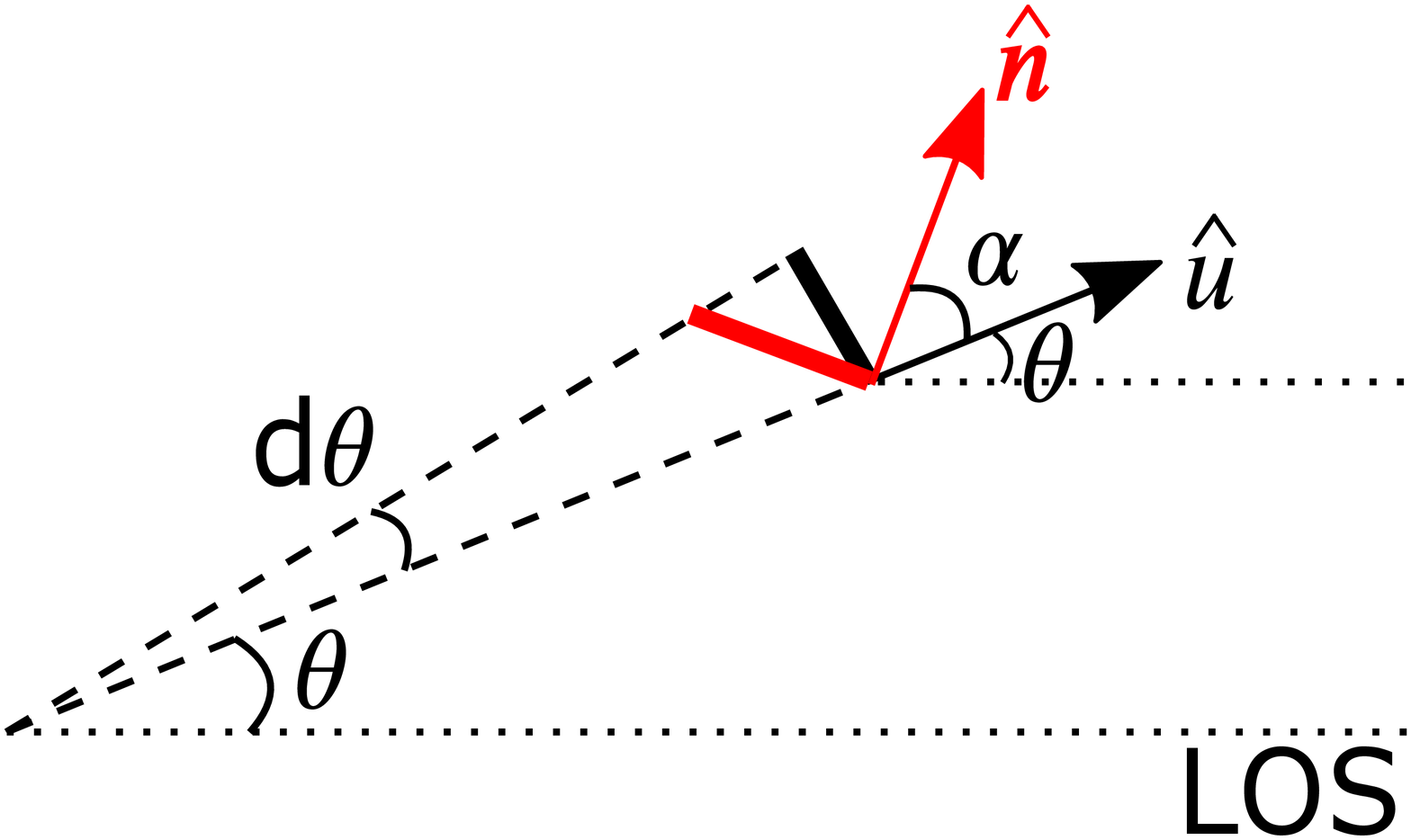} 
\caption{A small emitting patch (thick black line) of area $A$ moving in the $\hat{\b{u}}$ direction, which is also the normal direction of the patch. The angle between the velocity vector $\hat{\b{u}}$ and the observer's LOS is $\theta$. Since the patch is moving near the speed of light, the equal-arrival-time surface (thick red line, $\hat{\b{n}}$ being its normal direction) is tilted from the orientation of the patch in the lab frame by an angle $\alpha$, given by eq. (\ref{eq:EATS}). The apparent projected size of the patch seen by the observer is given by $A\cos(\alpha+\theta)/\cos\alpha$.
}\label{fig:inclination}
\end{figure}

We consider a small emitting patch of area $A$ moving in the direction normal to the surface at speed $\beta c$, and the observer's LOS is at an angle $\theta$ with respect to the velocity vector $\b{u}$ (or the normal) of the surface. In this section, we show that the apparent area of the surface is $A|\cos\theta'|$, i.e. the flux density received from the observer is given by
\begin{equation}
\label{eq:apparent_area}
    F_\nu = I_\nu A|\cos\theta'|/D^2,
\end{equation}
where $I_\nu$ is the specific intensity at the surface in the lab frame, $D$ is the distance to the surface, and $\theta'$ is the angle between the LOS and the surface normal in the comoving frame.

The surface is small so that it can be considered as a patch on a sphere that is expanding in the radial direction (the center of the sphere may not be coincident with the center of explosion). We consider $0<\theta<\pi/2$ only but the result is applicable for $\theta>\pi/2$ as well. Emission from different parts of the surface arrive at the observer at different times. At observer's time $\tau$, the equal-arrival-time (at $\tau$) surface are at radius
\begin{equation}
    r_{\rm eq}(\theta) = \beta c \tau/(1-\beta\cos\theta).
\end{equation}
We see that the angle between the velocity vector (the radial direction) and the normal vector of the equal arrival time surface $\alpha$ is given by
\begin{equation}
\label{eq:EATS}
    \tan\alpha = {1\over r_{\rm eq}} \left|{\partial r_{\rm eq}(\theta)\over \partial \theta}\right| = {\beta\sin\theta \over 1-\beta\cos\theta}.
\end{equation}
Thus, the physical surface area $A$ is stretched to a larger size of $A/\cos\alpha$ by the arrival time effects. With respect to the LOS, the equal-arrival-time surface is inclined at an angle $\alpha+\theta$, so the observer sees a solid angle spanned by the emitting patch $D^{-2}A\cos(\alpha+\theta)/\cos\alpha$ ($D$ being the distance). Therefore, the apparent projected size is given by
\begin{equation}
    {A\cos(\alpha+\theta)\over \cos\alpha} = A(\cos\theta - \tan\alpha \sin\theta) = A{\cos\theta -\beta \over 1-\beta \cos\theta}. 
\end{equation}
The multiplication factor is equal to $\cos\theta' = (\cos\theta -\beta)/(1-\beta \cos\theta)$ as given by Lorentz transformation of null rays. We note that in the optically thin limit, the intensity is proportional to the optical depth of the shell $\tau_\nu\propto N_e/(A|\cos\theta'|)$ as calculated in the comoving frame, so the observed flux density is independent of the orientation of the surface, in agreement with our physical intuition.

\section{Proper Motion of the Flux Centroid}
\begin{figure}
  \centering
\includegraphics[width = 0.35\textwidth]{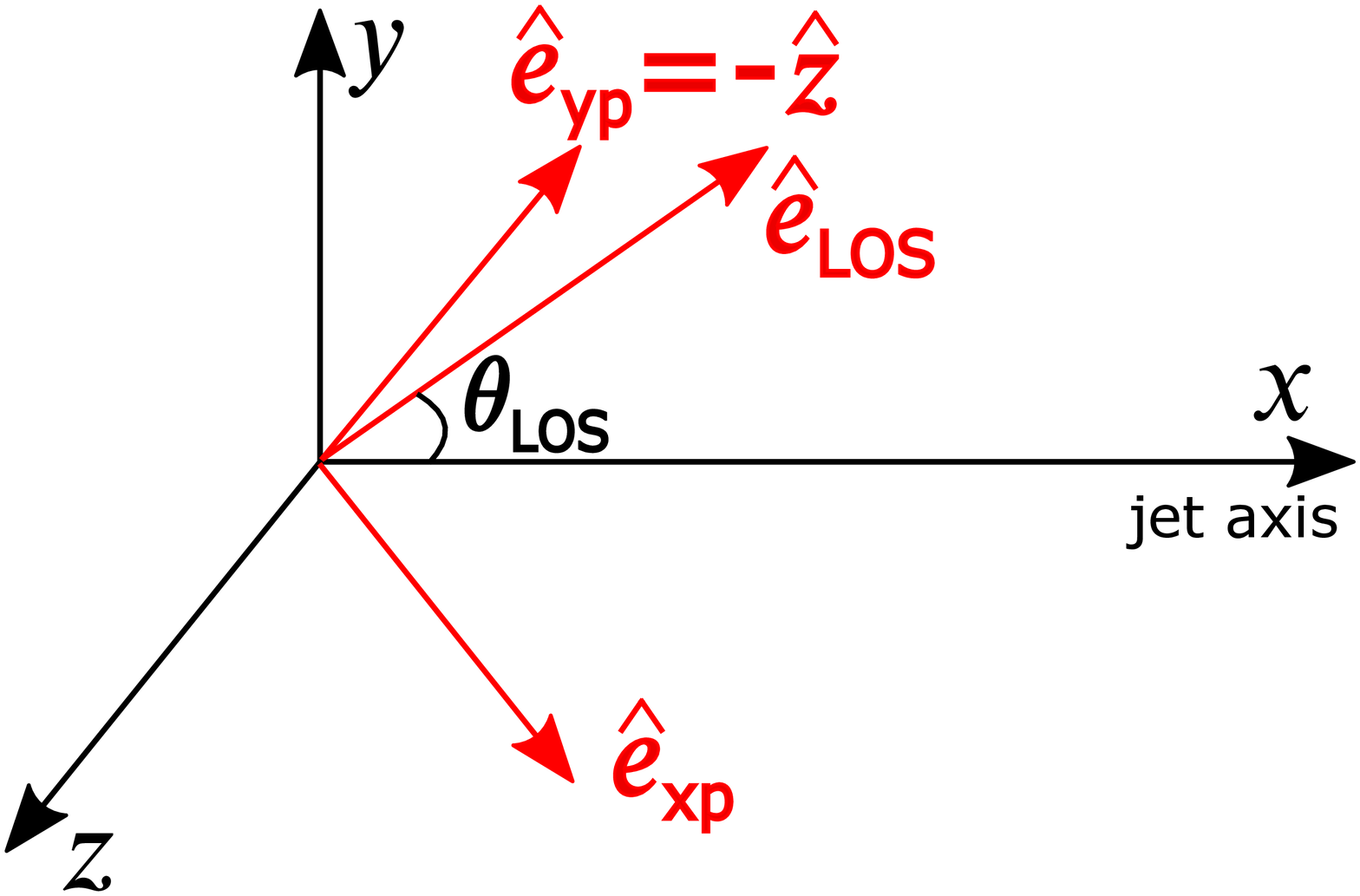} 
\caption{The relative positions of two coordinate systems. Black arrows show the Cartesian base vectors of the coordinate system we have used to describe the jet hydrodynamics. The jet axis is aligned with $\hat{\b{x}}$, and the observer's LOS is along the direction of $\hat{\b{e}}_{\rm LOS}$, which is at an angle $\theta_{\rm LOS}$ from $\hat{\b{x}}$. The base vectors of the plane of the sky are $\hat{\b{e}}_{\rm xp}$ and $\hat{\b{e}}_{\rm yp}$, and we choose $\hat{\b{e}}_{\rm xp}$ to be in the x-y plane and $\hat{\b{e}}_{\rm yp}$ to be along $-\hat{\b{z}}$. 
}\label{fig:projection}
\end{figure}

We project the position of a fluid element $\b{r}$ expressed in the coordinate centered on the explosion center as $(x, y, z) = r(\cos\theta, \sin\theta \cos\phi, \sin\theta \sin \phi)$ onto the plane of the sky. The jet axis is aligned with $\hat{\b{x}}$, and the observer's LOS is along the direction of $\hat{\b{e}}_{\rm LOS}$, which is at an angle $\theta_{\rm LOS}$ from $\hat{\b{x}}$. The base vectors of the plane of the sky are $\hat{\b{e}}_{\rm xp}$ and $\hat{\b{e}}_{\rm yp}$, and we choose $\hat{\b{e}}_{\rm xp}$ to be in the x-y plane (so the jet axis projected on the plane of the sky is along $\hat{\b{e}}_{\rm xp}$), and $\hat{\b{e}}_{\rm yp}$ is along $-\hat{\b{z}}$. The geometry is shown in Fig. \ref{fig:projection}. The projected position of $\b{r}$ is given by $(x_{\rm p}, y_{\rm p}) = (\b{r}\cdot \hat{\b{e}}_{\rm xp}, \b{r}\cdot \hat{\b{e}}_{\rm yp})$, i.e.,
\begin{equation}
\begin{split}
    x_{\rm p} &= r(\sin\theta_{\rm LOS} \cos\theta - \cos\theta_{\rm LOS}\sin\theta \cos\phi),\\
    y_{\rm p} &= -r\sin\theta \sin\phi.
\end{split}
\end{equation}

The observer is at an angular-diameter distance $D_{\rm A}$ away from the host galaxy, so the angular position of $\b{r}$ is $(\theta_{\rm xp}, \theta_{\rm yp}) = (x_{\rm p}/D_{\rm A}, y_{\rm p}/D_{\rm A})$. At each given observer's time $\tau$, we calculate the intensity $I_\nu$ from the surface of each fluid element on the equal-arrival-time surface as a function of projected position $(\theta_{\rm xp}, \theta_{\rm yp})$. Since the jet is symmetric in the $\hat{\b{e}}_{\rm yp}$ direction, the flux centroid is described by the intensity-weighted mean position $\bar{\theta}_{\rm xp}$, given by
\begin{equation}
    \bar{\theta}_{\rm xp} = {\iint \theta_{\rm xp} I_{\nu}(\theta_{\rm xp}, \theta_{\rm yp}) \d \theta_{\rm xp} \d \theta_{\rm yp} \over
    \iint I_{\nu} \d \theta_{\rm xp} \d \theta_{\rm yp}}.
\end{equation}
The apparent angular speed (or proper motion) of the flux centroid between two epochs $t_{\rm obs,1}$ and $t_{\rm obs,2}$ is given by
\begin{equation}
    v_{\theta} = {\bar{\theta}_{\rm xp}(t_{\rm obs,2}) - \bar{\theta}_{\rm xp}(t_{\rm obs,1}) \over t_{\rm obs,2} - t_{\rm obs,1}},
\end{equation}
in units of radian per second.



\section{Narrow Power-law Jet}
\begin{figure}
  \centering
\includegraphics[width = 0.45\textwidth, keepaspectratio]{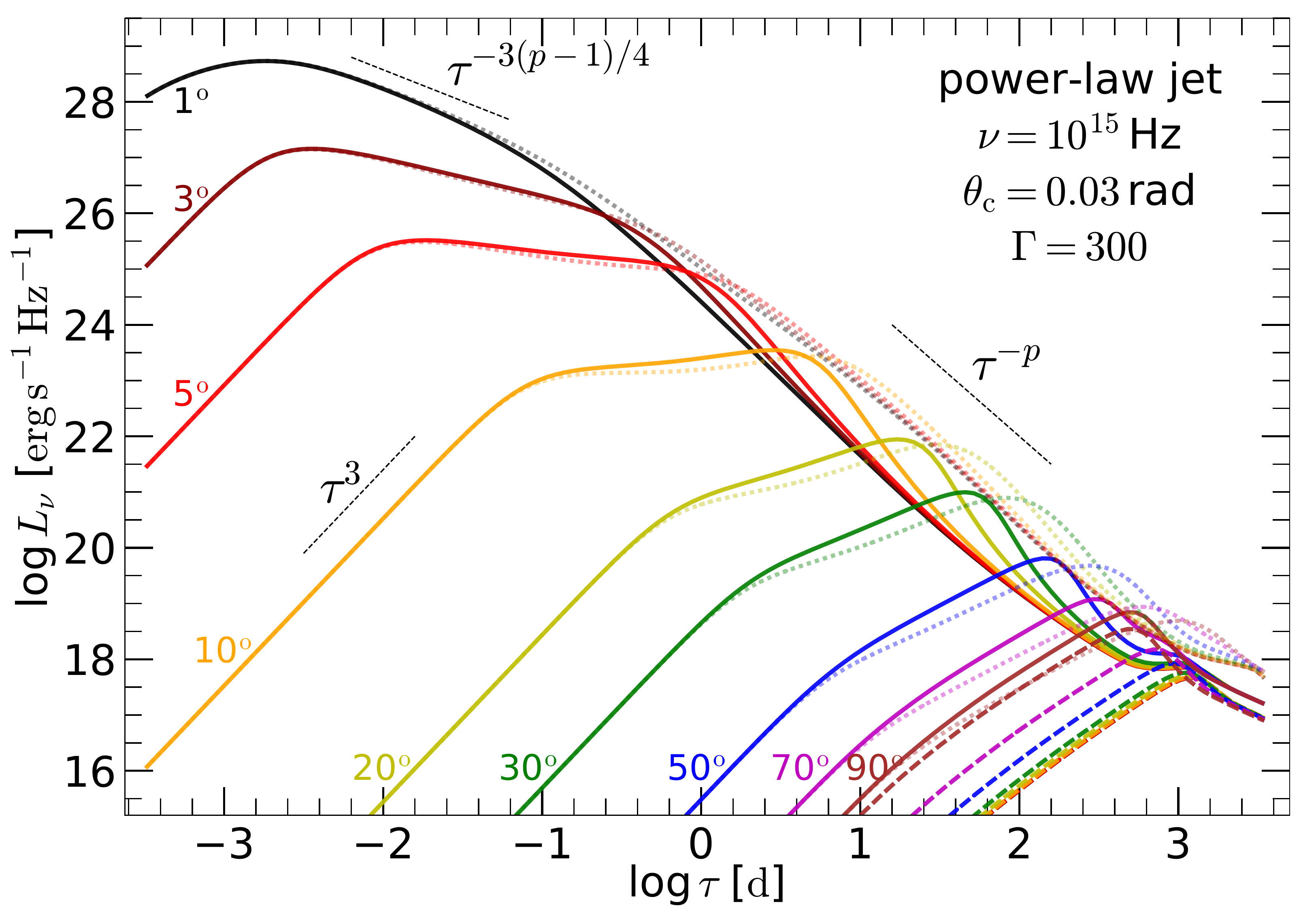}
\includegraphics[width = 0.45\textwidth, keepaspectratio]{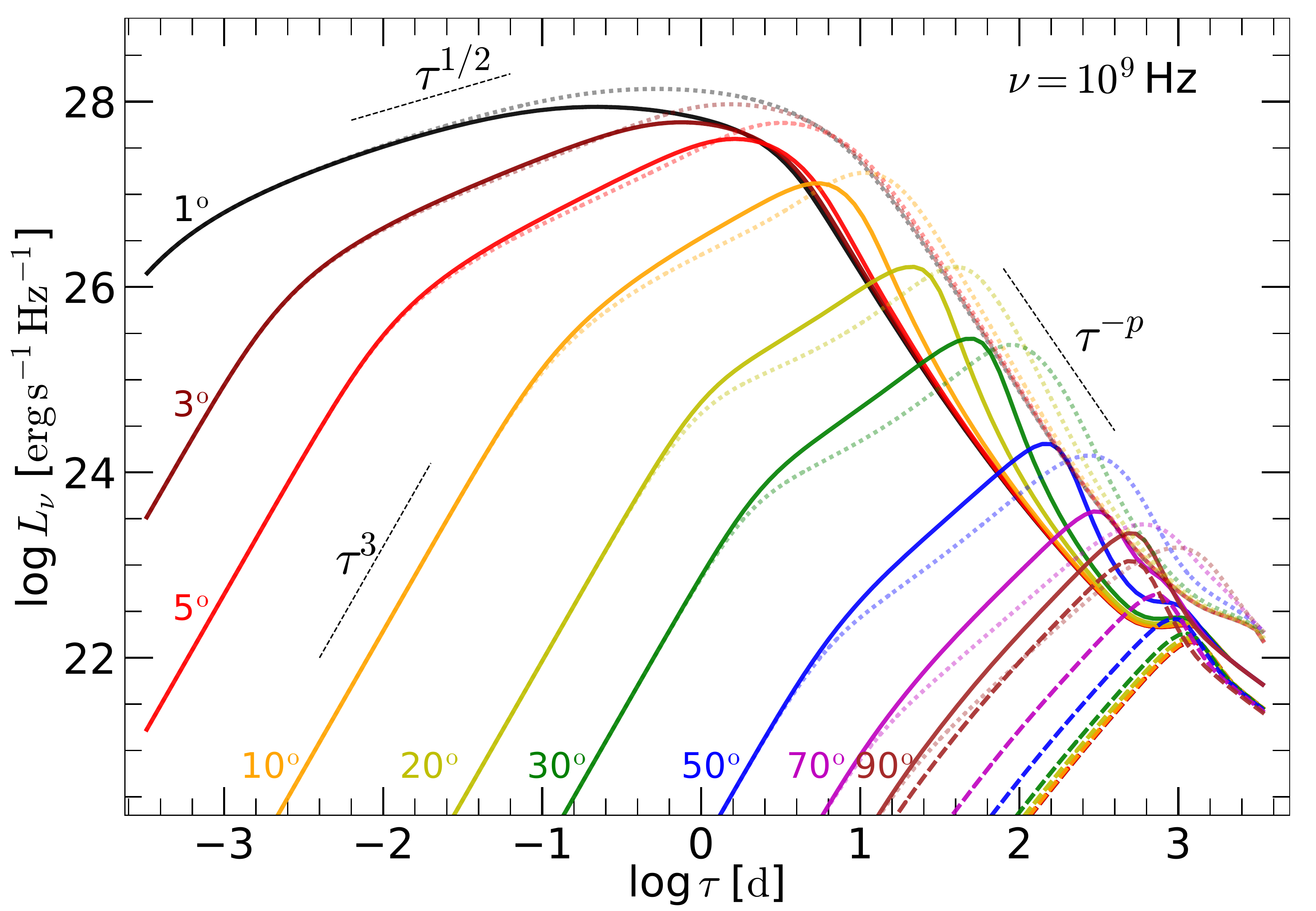}
\caption{Lightcurves for a narrowly collimated power-law jet at $\nu=10^{15}\,$Hz (left panel, $\num<\nu<\nuc$) and $10^9\,$Hz (right panel, $\nu<\num$ in $L_\nu\propto \tau^{1/2}$ phase but later on $\num<\nu<\nuc$) from different viewing angles marked along each line. The initial conditions differ from that in Fig. \ref{fig:PLjet_LC} only in $\theta_{\rm c}=0.03$ and $u_{\rm 0,max}=300$. The solid lines are for \textit{total} flux (including both forward and counter jets), the dashed lines are for contribution from the counter jet only. The faint dotted lines are for the total flux from the same jet but without lateral expansion. Microphysical parameters for the forward shock are $\epse=0.1$, $\epsB=10^{-4}$, $p=2.5$.
}
\label{fig:PLjet_LC_narrow}
\end{figure}

\begin{figure*}
  \centering
\includegraphics[width = 0.8\textwidth, keepaspectratio]{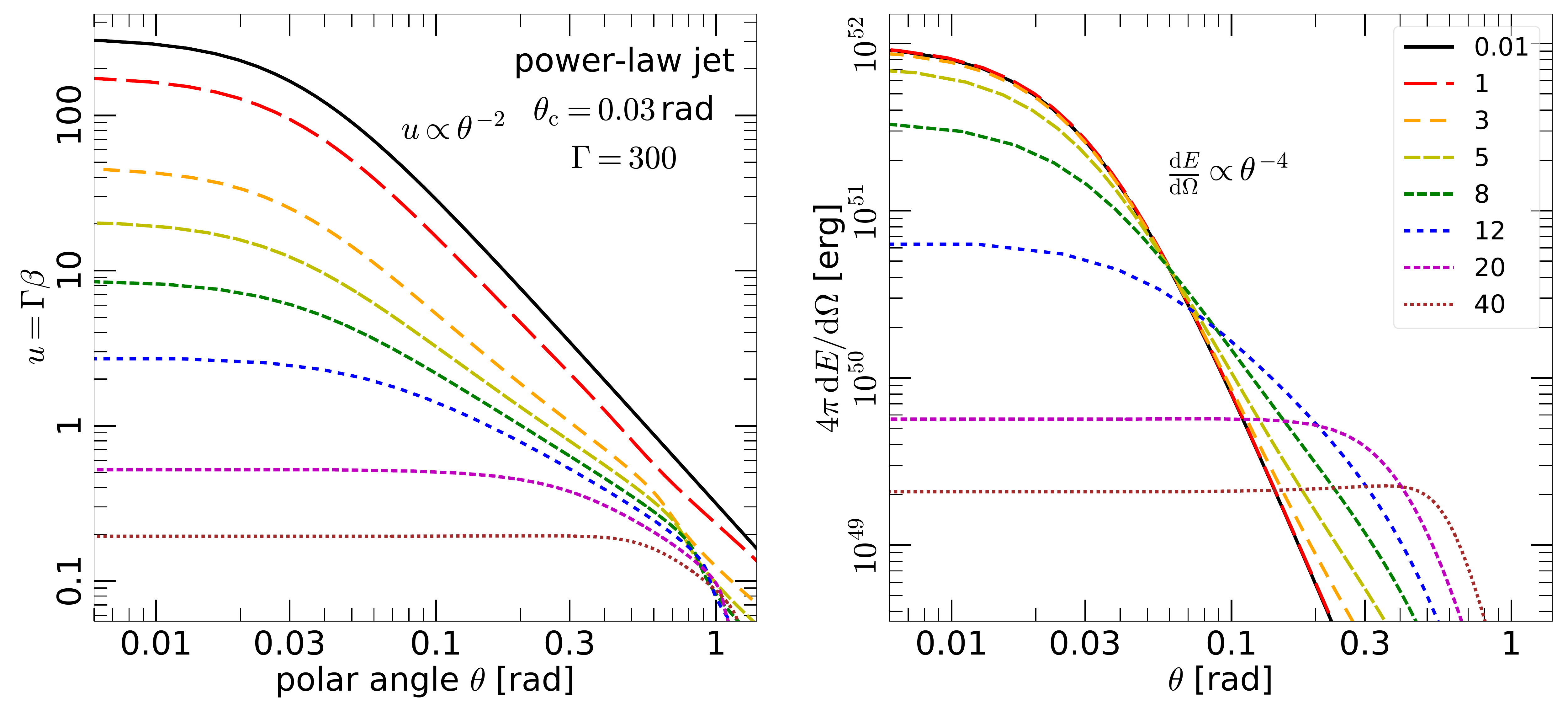}
\caption{Evolution of angular structures $u(\theta)$ (left panel) and $\d E/\d \Omega(\theta)$ (right panel) for a power-law jet with peak isotropic energy $E_{\rm iso}=10^{52}\rm\,erg$, core size $\theta_{\rm c}=0.03\,$rad, peak four-velocity $u_{0,\rm max}=300$, energy structure index $q=4$, four-velocity structure index $s=2$, and uniform ambient medium density $n=10^{-2}\rm\,cm^{-3}$. The time for each snapshot is shown in the legend of the right panel in units of the lab-frame deceleration time $t_{\rm dec}=45\,$d as defined in eq. (\ref{eq:44}), from $t/t_{\rm dec}=0.01$ to 40.
}
\label{fig:PLjet_narrow}
\end{figure*}

\begin{figure*}
  \centering
\includegraphics[width = 0.9\textwidth, keepaspectratio]{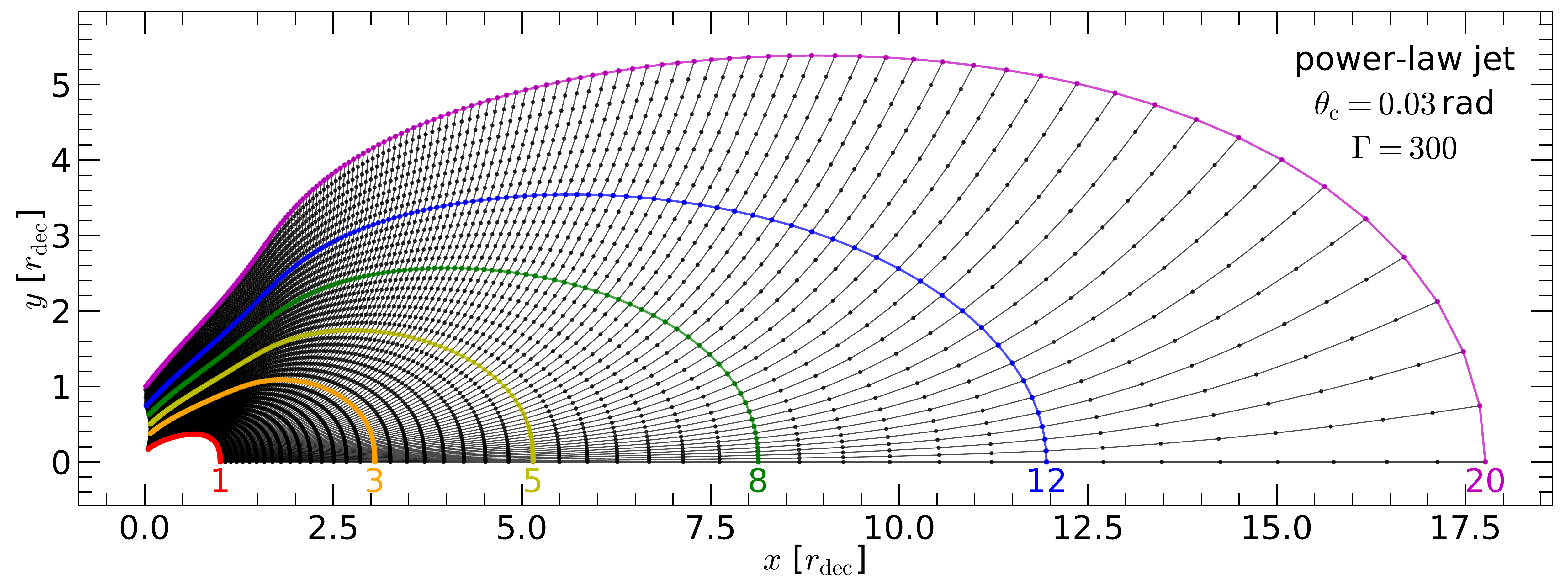}
\caption{Trajectories of the grid points from $t/t_{\rm dec}=1$ to 20 (black curves) for a power-law jet as described in Fig. \ref{fig:PLjet_narrow}. We highlight the positions of the jet surface at a number of epochs from $t/t_{\rm dec}=1$ (innermost red curve), 3, 5, 8, 12 to 20 (outermost magenta curve). The physical units are $r_{\rm dec}=3.8\times10^{-2}\,$pc and $t_{\rm dec}=45\,$d as defined in eq. (\ref{eq:44}). 
}\label{fig:PLjet_2D_narrow}
\end{figure*}

We also show the results from a narrow power-law jet case with $\thec=0.03\,$rad and $u_{\rm 0,max}=300$ in Figs. \ref{fig:PLjet_LC_narrow}, \ref{fig:PLjet_narrow}, and \ref{fig:PLjet_2D_narrow}. Other parameters are shown in the caption of Fig. \ref{fig:PLjet_narrow}.

\label{lastpage}
\end{document}